\documentclass[useAMS,usenatbib]{mn2e}

\newcommand{\OIII}{\mbox{[O\,\textsc{III}]}}
\newcommand{\OIIIa}{\mbox{[O\,\textsc{III}] $\lambda$5007}}
\newcommand{\OIIIb}{\mbox{[O\,\textsc{III}] $\lambda$4959}}
\newcommand{\OIIIc}{\mbox{[O\,\textsc{III}] $\lambda$4363}}
\newcommand{\NII}{\mbox{[N\,\textsc{II}]}}
\newcommand{\NIIa}{\mbox{[N\,\textsc{II}] $\lambda$6583}}
\newcommand{\NIIb}{\mbox{[N\,\textsc{II}] $\lambda$6548}}
\newcommand{\OI}{\mbox{[O\,\textsc{I}]}}
\newcommand{\OII}{\mbox{[O\,\textsc{II}]}}
\newcommand{\SII}{\mbox{[S\,\textsc{II}]}}
\newcommand{\SIIa}{\mbox{[S\,\textsc{II}] $\lambda$6717}}
\newcommand{\SIIb}{\mbox{[S\,\textsc{II}] $\lambda$6731}}
\newcommand{\HII}{\mbox{H\,\textsc{II}}}
\newcommand{\HeII}{\mbox{He\,\textsc{II}}}
\newcommand{\NeIII}{\mbox{[Ne\,\textsc{III}]}}
\newcommand{\NeV}{\mbox{[Ne\,\textsc{V}]}}
\newcommand{\NIIHa}{\NII/H$\alpha$}
\newcommand{\SIIHa}{\SII/H$\alpha$}
\newcommand{\OIIIHb}{\OIII/H$\beta$}
\newcommand{\Ha}{H$\alpha$}
\newcommand{\Hb}{H$\beta$}
\newcommand{\Hg}{H$\gamma$}

\usepackage{epsfig,graphicx}
\usepackage{amsmath} 
\usepackage{gensymb}	
\usepackage[3D]{movie15}
\usepackage{hyperref}

\usepackage{aas_macros}
\begin{document}

\title[J224024.1--092748: Quasar ionization echo]{The ``Green Bean'' Galaxy SDSS J224024.1--092748: Unravelling the emission signature of a quasar ionization echo}

\author[R. L. Davies, M. Schirmer, J. E. H. Turner]{Rebecca L. Davies$^{1,2}$\thanks{E-mail:
Rebecca.Davies@anu.edu.au (RLD); mschirme@gemini.edu (MS)}, Mischa Schirmer$^{3}$,
James\ E.\ H.\ Turner$^{3}$ \\
$^{1}$Research School of Astronomy and Astrophysics, Australian National University, Cotter Road, Weston, ACT 2611, Australia \\
$^{2}$Australian Astronomical Observatory, P.O. Box 915, North Ryde,
NSW 1670, Australia \\
$^{3}$Gemini Observatory, Casilla 603, La Serena, Chile.}

\maketitle

\begin{abstract}
``Green Bean'' Galaxies (GBs) are the most \OIII-luminous type-2 active galactic nuclei (AGN) at $z\sim0.3$. However, their infrared luminosities reveal AGN in very low activity states, indicating that their gas reservoirs must be ionized by photons from a recent high activity episode - we are observing quasar ionization echoes. We use integral field spectroscopy from the Gemini Multi-Object Spectrograph to analyse the 3D kinematics, ionization state, temperature and density of ionized gas in the GB SDSS \mbox{J224024.1--092748}. We model the emission line spectrum of each spaxel as a superposition of up to three Gaussian components and analyse the physical properties of each component individually. Two narrow components, tracing the velocity fields of the disc and an ionized gas cloud, are superimposed over the majority of the galaxy. Fast shocks produce hot (\mbox{$T_e \geq$ 20,000 K}), dense (\mbox{$n_e \geq 100 \rm \, cm^{-3}$}), turbulent (\mbox{$\sigma \geq 600 \rm \, km \, s^{-1}$}), \OIII-bright regions with enhanced \NIIHa\ and \SIIHa\ ratios. The most prominent such spot is consistent with a radio jet shock-heating the interstellar medium. However, the AGN is still responsible for \mbox{$\ga$ 82 per cent} of the galaxy's total \OIII\ luminosity, strengthening the case for previous quasar activity. The ionized gas cloud has a strong kinematic link to the central AGN and is co-rotating with the main body of the galaxy, suggesting that it may be the remnant of a quasar-driven outflow. Our analysis of \mbox{J224024.1--092748} indicates that GBs provide a unique fossil record of the transformation from the most luminous quasars to weak AGN. 
\end{abstract}


\section{Introduction}\label{Sec:intro}
Extended ionized gas reservoirs around active galactic nuclei (AGN) are valuable probes of the mechanisms driving rapid changes in AGN accretion rates. Kiloparsec-scale regions of highly ionized gas (narrow line regions; NLRs) are excited by energetic outflows, driven by either radiation pressure (in luminous AGNs with high accretion rates; see e.g. \citealt{Alexander10}) or radio cores/jets (in radio-loud systems; see e.g. \citealt{Rosario10, Mullaney13}). The most luminous AGN can ionize gas to distances of tens of kiloparsecs, forming extended emission line regions (EELRs). EELRs have significantly lower surface brightnesses than typical AGN NLRs, but require substantially more energy to ionize due to their large volumes (see \citealt{Stockton06} for a review). For this reason, EELRs have been observed almost exclusively around radio-loud quasars \citep{Fu09}. 

The small subset of EELRs found around galaxies with weak AGN are ionization echoes and provide fossil records of the rapid shutdown of luminous quasars. The long recombination timescale in low density EELR gas allows it to remain highly ionized for tens of thousands of years after the quasar phase of the central engine has ceased. Hanny's Voorwerp, a large and highly ionized gas cloud associated with the galaxy \mbox{IC 2497} (\mbox{z $\approx$ 0.05}), was discovered serendipitously in the Galaxy Zoo survey \citep{Lintott09, Schawinski10, Keel12b}. The lack of current AGN activity in \mbox{IC 2497} indicates that the luminosity of the AGN must have decreased by at least 2-3 orders of magnitude in less than the light crossing time from the nucleus to Hanny's Voorwerp ($10^{4-5}$ years, see \citealt{Lintott09, Schawinski10, Keel12b}). A targeted Galaxy Zoo follow-up search revealed 8 further AGN ionization echoes at \mbox{z $<$ 0.1} with similar properties to Hanny's Voorwerp \citep{Keel14}. These objects provide unique insights into both the life cycle of AGNs and the physical processes driving rapid changes in AGN accretion rate.  

The descendants of the most luminous quasars are very rare and can therefore only be found by increasing the search volume to higher redshifts. The prototypical quasar ionization echo SDSS J224024.1--092748 (J2240 hereafter), at a redshift of \mbox{z = 0.326}, was discovered serendipitously in a Canada France Hawaii Telescope (CFHT) image by \citet{Schirmer13}. A follow-up search using the Sloan Digital Sky Survey (SDSS) Data Release 8 archive produced a sample of 29 `Green Bean' galaxies (GBs) across two redshift ranges (\mbox{0.19 $<$ z $<$ 0.35} and \mbox{0.39 $<$ z $<$ 0.69}). The GBs are defined by their extreme \OIII\ $\lambda$5007 fluxes ($r\sim18$ mag at \mbox{z $\sim$ 0.25}) and large angular extents (several arcseconds; \citealt{Schirmer13}). GBs occupy a volume in color-magnitude-size space that is populated by about 95 per cent spurious detections which must be removed by visual inspection.

GBs are the most luminous AGN at their cosmic time. Their EELRs span \mbox{$20-40$ kpc} in diameter and have typical \OIII\ luminosities of \mbox{${\rm log}(L_{[OIII]}$ / erg $s^{-1}$) = 42.7 - 43.6}, making them $\sim10$ times more luminous than the 1300 reference Type-2 quasars at similar redshifts selected from \citet{Reyes08} and \citet{Mullaney13}. Only half of the known GBs have radio luminosities larger than \mbox{$10^{23}$ W$\,\rm{Hz}^{-1}$} compared to 98 per cent of all type-2 AGN at similar redshifts \citep{Mullaney13}, demonstrating that their EELRs must be excited largely by radiation pressure. The EELRs of GBs lie well above the empirical size-luminosity relation \citep[see e.g.][]{Liu13}, indicating that their extreme luminosities may be a by-product of larger-than-average gas reservoirs \citep[see e.g.][]{Hainline13}. 

The most unusual property of GBs is that their 22$\mu$m WISE IR luminosities ($L_{22 \mu m}$) are $5-50$ times lower than expected from $L_{\rm [O~III]}$ \citep{Schirmer13}. Both $L_{22 \mu m}$ and $L_{\rm [O~III]}$ probe AGN activity \citep{Asmus11, Ichikawa12}. The infrared (IR) flux in active galaxies is emitted by heated dust within $10-100$ pc of the SMBH, whereas the \OIII\ emission originates from the EELR on kiloparsec scales. Therefore, the discrepancy between $L_{22 \mu m}$ and $L_{\rm [O~III]}$ may be best explained by a recent AGN shutdown, which due to the long gas recombination timescale has not yet propagated through the EELR. Another explanation is that the dusty torus emits disproportionally less IR light at a given AGN bolometric luminosity, due to e.g. structural differences in the torus (Hai Fu, private communication). However, it is not conceivable that the SDSS color selection adopted by \citet{Schirmer13} could bias the luminosity of our sample enough to produce such an effect. Processes involving the orbital decay of binary AGN or the coalescence of two SMBHs cannot be responsible for the discrepancy, as the X-ray emissions involved are too low and short-lived \citep{Hopkins10, Tanaka10, Khan12}. 

Preliminary analysis of new Chandra data for 9 GBs reveals X-ray counts much lower than expected from their mid-IR luminosities in the limiting case of Compton-thick (${\rm log} [N_{\rm H}] > 24.2$ cm$^{-2}$) absorbers. Either these AGN are deeply buried, or we are witnessing the descendants of luminous quasars $10^{2-4}$ years (i.e. much less than the light crossing time of the EELR) after the shutdown of their quasar phases. The large angular extent of the EELRs allow them to retain a memory of the past X-ray histories of the AGNs, spanning up to $\sim10^5$ years. Our Chandra observations will be described in detail in a future paper.

Using photoionization models and resolved 3D optical spectroscopy, it is possible to extrapolate the incident X-ray flux at a given distance from the nucleus to infer the nuclear X-ray luminosity for different times in the past. The EELRs of GBs are sufficiently bright and luminous that light curve reconstruction is possible. However, such reconstruction requires a robust estimate of the contribution of shock ionization, and a detailed understanding of the formation, composition and physical state of the gas components in the EELR. In this paper we present deep optical integral field spectroscopy of the GB prototype J2240, spanning the 330nm$-$680nm optical rest-frame wavelength range.  We explain our observations, data reduction procedure and data processing techniques in Section \ref{sec:obs}, and decompose the gas into different kinematic components in Section \ref{sec:initial_analysis}. We present a detailed 3D spectroscopic analysis of the velocity field, diagnostic line ratios, ionization parameter, and electron temperature and density in Section \ref{sec:3D_analysis} and discuss the origin of the EELR in Section \ref{sec:discussion}. We summarise our results and present our conclusions in Section \ref{sec:conclusions}.   

\section{Observation and Data Processing}
\label{sec:obs}

\begin{table}
\begin{tabular}{c|c|c|c} \hline \hline
Channel               & Blue       & Green      & Red            \\ \hline
Instrument			  & GMOS-N		& GMOS-S	 & GMOS-N			\\
Grating/Filter        & B600 + g   & B600 + r   & R831 + z + CaT \\
Date obtained         & 2012-10-08 & 2012-08-27 & 2012-10-18     \\
$\lambda_{obs}$ (nm)  & 429 - 542  & 559 - 688  & 847 - 905      \\
$\lambda_{rest}$ (nm) & 323 - 409  & 421 - 520  & 638 - 683      \\
R                     & 2700       & 2700       & 7030           \\
Exposure time (s)     & 4 x 1800   & 6 x 1500   & 4 x 1600       \\
Mean airmass          & 1.17       & 1.10       & 1.19          \\ \hline
\end{tabular}
\caption{Characteristics of GMOS integral field unit (IFU) observations.}
\label{Table:obs}
\end{table}

\subsection{Gemini observations}
We obtained deep Gemini Multi-Object Spectrograph (GMOS) observations of J2240 under observing programmes \mbox{GS-2012B-Q-26} and \mbox{GN-2012B-Q-226} (PI: Schirmer). The two GMOS instruments (GMOS-S and GMOS-N) have four observing modes: imaging, long-slit spectroscopy, multi-object spectroscopy and integral field spectroscopy \citep{Allington-Smith02, Hook04}. Our integral field unit (IFU) observations cover three spectral ranges: blue (\mbox{B600 + g}), \mbox{$\lambda_{obs}$ = 429 - 542 nm}, \mbox{R $\approx$ 2700}), green (\mbox{B600 + r}, \mbox{$\lambda_{obs}$ = 559 - 688 nm}, \mbox{R $\approx$ 2700}) and red (\mbox{R831 + z + CaT}, \mbox{$\lambda_{obs}$ = 847 - 905 nm}, \mbox{R $\approx$ 7030}). The wavelength coverage, spectral resolution, exposure times and dates and mean airmasses of our observations are summarised in Table \ref{Table:obs}. The observations were carried out in IQ 70 conditions, corresponding to a Point Spread Function (PSF) with a Full Width at Half Maximum (FWHM) of approximately \mbox{0.5 - 0.6 arcsec} (5-6 pixels) across all three spectral channels. The angular extent of the target over which we conduct spectroscopic analysis is \mbox{2.5$\arcsec$ $\times$ 5$\arcsec$}, within which there are 32-50 separate resolution elements.

We observed spectrophotometric standard stars for flux and wavelength calibration of all three spectral channels: \mbox{BD +28 4211} on 2012-10-05, \mbox{LTT 9239} on 2012-10-31 and \mbox{EG 131} on 2012-08-29, and Wolf 1346 on 2014-04-03 for the blue, green and red channels respectively. We also obtained the other standard GMOS baseline calibrations including night-time flats, twilight spectra, day-time arcs \& biases.	

\subsection{Data reduction}
\label{subsec:data_reduction}
Our data were reduced with the Ureka\footnote{\href{http://ssb.stsci.edu/ureka}{http://ssb.stsci.edu/ureka}}\,1.0 version of IRAF 2.16 \citep{Tody86, Tody93}, using our own modified version of Gemini's GMOS IRAF package, which we refer to as \emph{ifudrgmos}, along with our \emph{PyFU} datacube manipulation scripts. Our modifications are described in the Appendix. Both our \emph{ifudrgmos} and \emph{PyFU} packages are available to other investigators through the Gemini Data Reduction User Forum\footnote{\href{http://drforum.gemini.edu}{http://drforum.gemini.edu}}. For our GS-2012B-Q-26 and GN-2012B-Q-226 data we used revisions 79 and 114, respectively, of the Subversion repository containing these packages, which are functionally the same as the versions on the forum (since the numbering tracks changes to multiple packages).

Our top-level processing sequence -- based on the standard GMOS data reduction -- was as follows: bias and overscan subtraction; creation and association of bad column masks; interpolation over bad columns using continuum fits; tracing of fibre spectra and gap regions using flat field exposures from the calibration unit, GCal; modelling and subtraction of scattered light in scientific and flat exposures based on the count levels in gaps between blocks of fibres; identification of cosmic rays; cleaning of cosmic rays using interpolated replacement values; mosaicking of the GMOS detectors into a single array; extraction of fibre spectra to one dimension; wavelength calibration; normalization of GCal flats; division of scientific and standard star data by the flat field spectrum; rectification to linear wavelength coordinates; sky subtraction using the background field (separately for each IFU slit); integration over the IFU field to obtain standard-star spectra; flux calibration with reference to IRAF's tabulated data from \citet{Bessell99}, \citet{Hamuy94} and \citet{Massey88}; correction for sub-\AA\ spectral flexure using sky lines; resampling the spectra to 3-dimensional `data cubes'; spatial registration (by cross-correlation) and co-addition of the data cubes for individual exposures within each wavelength channel; and re-binning of the co-added data cubes in logarithmic wavelength increments, resulting in constant radial velocity increments (see Sect. 2.2.8). The last few processing steps were also repeated identically without sky subtraction, providing a measure of the final spectral resolution.

Variance and data quality arrays were propagated throughout the process, overhauling a number of limitations and inaccuracies in the existing GMOS code. As usual, random measurement errors were initially estimated from the bias-subtracted data values in electrons (assuming a Poisson distribution) and the nominal detector read noise. The calculations do not include systematics, nor contributions from the high-signal-to-noise flat field and standard exposures. At each step, the variance estimate from the previous one was propagated using the standard uncorrelated error calculations corresponding to the arithmetic operations performed on the main data array. When resampling, the variance arrays were interpolated in the same way as the scientific data, ignoring any minor smoothing and covariance introduced on small scales (which are zero in the ideal limit of sinc interpolation). The greater covariance introduced by a change in pixel scale when creating data cubes was dealt with overall using a simple scaling factor (see Section \ref{subsec:cube_reconstruction}), to avoid having to track redundant, higher-dimensional information. Separate data quality bit planes were kept for detector defects, chip gaps and cosmic rays, allowing each to be treated at the most appropriate processing stage and in the corresponding dimensions.

Differences in our data reduction from the standard GMOS procedure are described in the Appendix. Following publication, we will also provide an example CL script for this process (and the associated software improvements) through the Gemini forum.

\begin{figure*}
\centerline{\includegraphics[scale=1, clip = true, trim= 10 0 0 95]{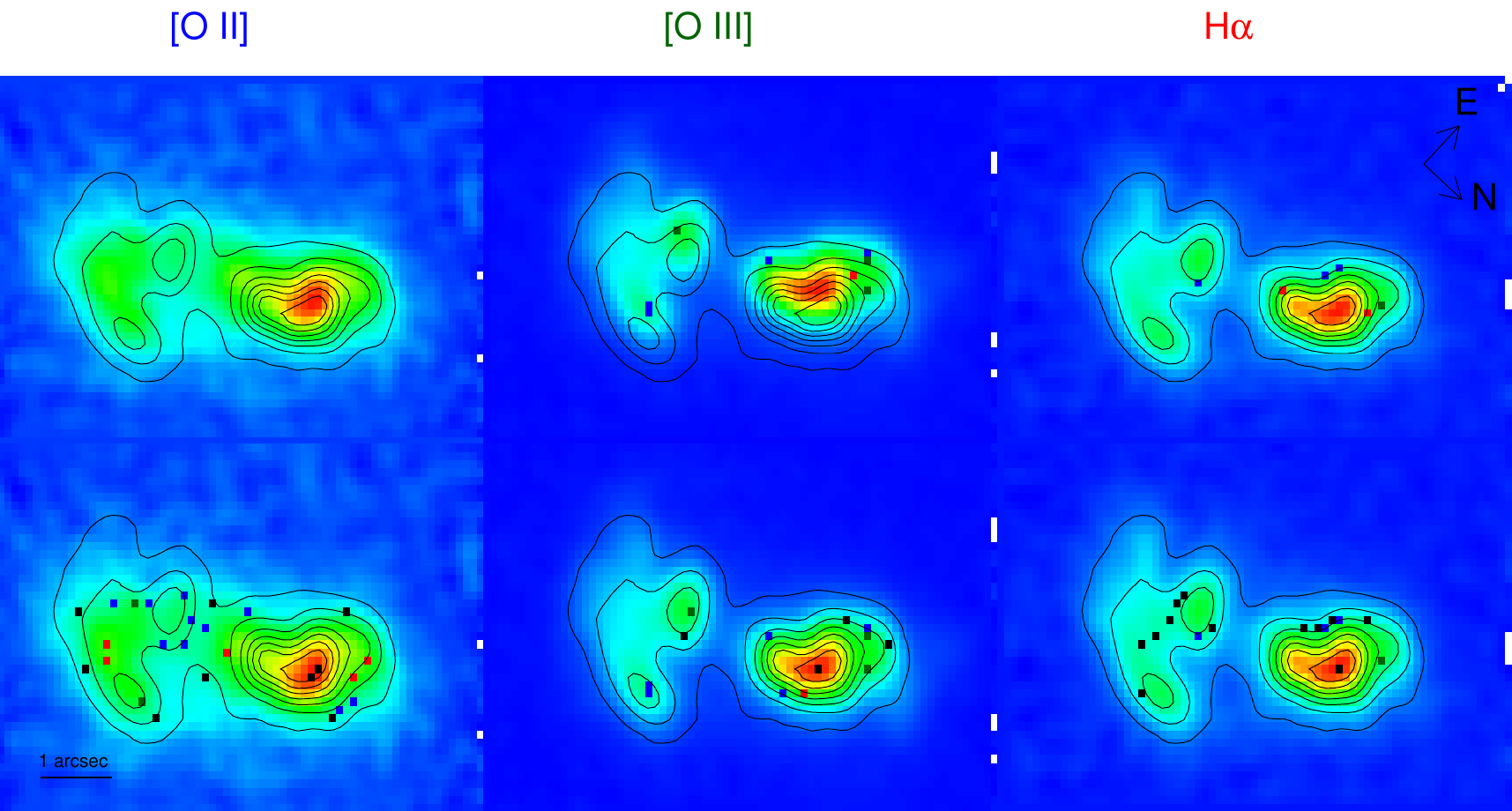}}
\caption{Top: Flux maps for the central 75 $\rm km \, s^{-1}$ of the \mbox{\OII\ $\lambda \lambda$3727, 3729}, \OIIIa\ and \mbox{\Ha\ $\lambda$6563} emission line profiles in the original data cubes. Contours trace the flux level in the \Ha\ slice. There is a relative offset of a few pixels in the \textit{y}-direction between the cubes. Bottom: The same flux maps extracted from the aligned data cubes. The spatial variation in the flux levels is very similar across all slices; indicating that the cubes are well aligned.}\label{Fig:alignment}
\end{figure*}

\subsection{Data cube alignment and continuum fitting}
We fit the stellar continuum and nebular emission spectra of each spaxel across all three spectral channels simultaneously (see Section \ref{subsubsec:schematic}), and therefore it is critical that the relative spatial alignment between the data cubes is accurate to within one pixel. We spatially align the data cubes by comparing flux maps for the central 75 $\rm km \, s^{-1}$ of the \mbox{\OII\ $\lambda \lambda$3727, 3729}, \OIIIa\ and \mbox{\Ha\  $\lambda$6563} emission line profiles (see Figure \ref{Fig:alignment}). Contours trace the flux level in the \Ha\ slice. The original data cubes (top row) have a relative offset of a few pixels in the \textit{y}-direction. However, this was accurately corrected for by eye using the spatial structure in the flux maps. 
 
The stellar continuum is modelled as a linear combination of stellar templates using the penalized pixel fitting routine (PPXF; \citealt{Cappellari04}). We adopt the high spectral resolution (\mbox{$\Delta \lambda$ = 0.3 \AA}) stellar evolutionary synthesis models of \citet{GonzalezDelgado05}, using isochrones from \citet{Bertelli94, Girardi00}, and \citet{Girardi02}. The templates cover 3 metallicities (Z = 0.019, 0.008, 0.004) and 74 ages (\mbox{4 Myr \textless\ t \textless\ 17 Gyr}) with a 0.06 log time sampling. The principal aim of the stellar population fitting is to correct for Balmer absorption produced by old stellar populations, and therefore the exact choice of stellar templates will have a minimal impact on our results. We subtract the fitted stellar continuum cubes from the full data cubes to produce emission line cubes.

\begin{figure*}
{\includegraphics[scale=0.95,clip = true, trim = 0 50 0 0]{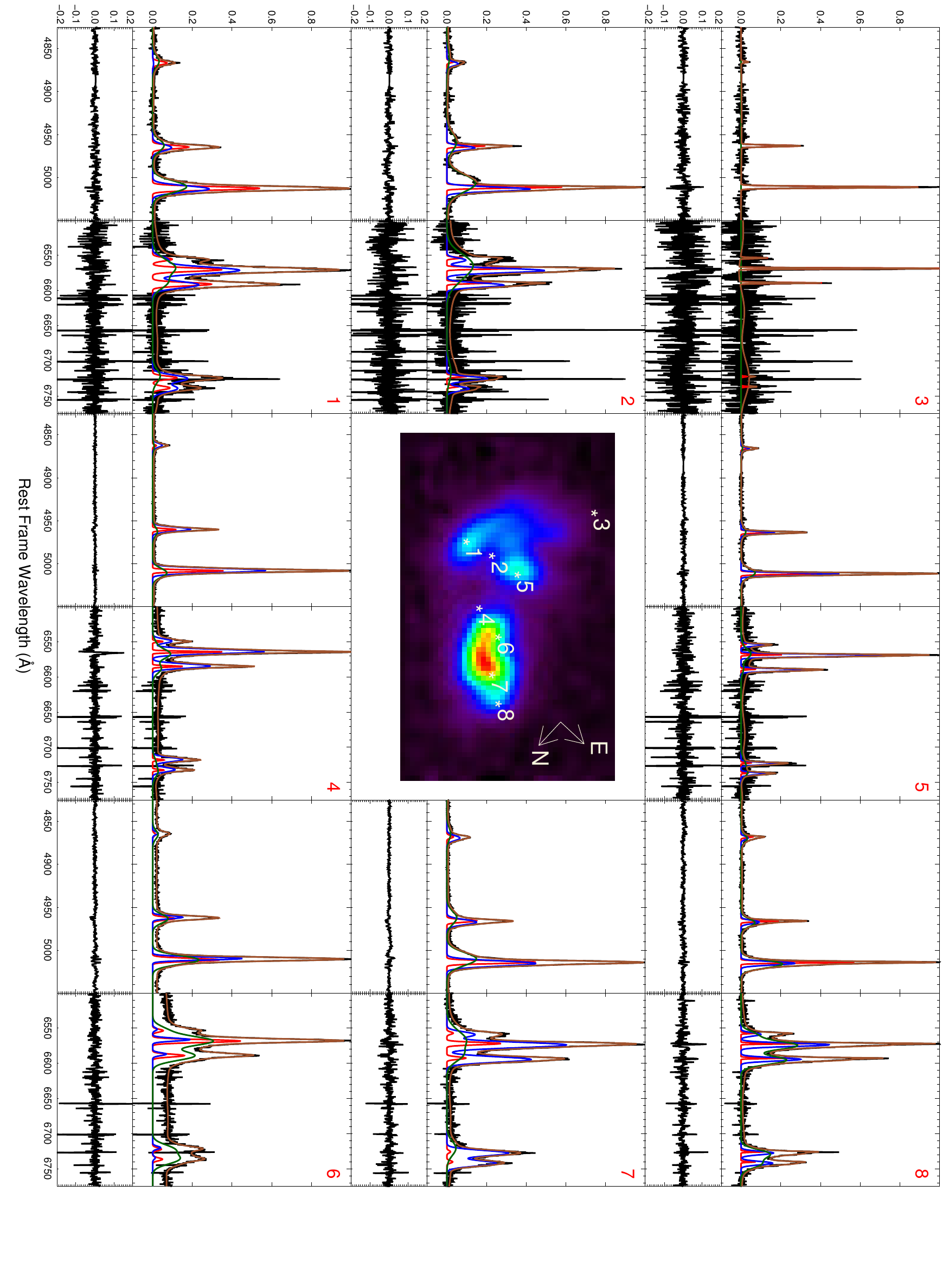}}
\caption{Schematic showing reduced spectra (black) and emission line fits (individual kinematic components shown in red, blue and green) as well as the total fits (brown) for eight different regions in J2240. Red numbers in the top right hand corner of each sub-plot correspond to the numbered locations in the central image of the galaxy. The two upper panels in each sub-plot show spectra zoomed in on the \Hb\ + \OIII\ complex (4800-5100 \AA; left hand side) and the \NII\ + \Ha\ and \SII\ complexes (6450-6800 \AA; right hand side). The two lower panels show the fitting residuals for each spectral region. The fluxes are normalized to the peaks of the \OIII\ and \Ha\ lines in each spectral range individually. The fitting residuals are on the order of a few percent, indicating that our parametric emission line fitting successfully reproduces the observed spectra in J2240.}
\label{Fig:schematic}
\end{figure*}

\subsection{Emission line fitting}
\label{subsubsec:schematic}
3D analysis of emission line fluxes and emission line ratios is one of the most powerful probes of the ionizing sources and kinematics of emission line galaxies. Emission line luminosities probe the total energy output of individual ionizing sources, whilst emission line ratios provide information regarding the temperature, density, metal abundance and excitation state of the line-emitting gas. The wavelength centroid and shape of the emission line profiles also encode detailed information about 3D kinematic structure. 

The emission line spectra of galaxies can be analysed either parametrically or non-parametrically, with advantages and disadvantages associated with both methods. Non-parametric emission line flux measurements are made by integrating the flux within specific wavelength bandpasses. The gas velocity is estimated by comparing the intrinsic wavelength of an emission line to the wavelength where the flux peaks, and the velocity dispersion is estimated using the FWHM of the emission line. The principal advantage of non-parametric analysis is that no assumption is made regarding the shapes of the line profiles. This allows for accurate measurement of the total flux in isolated emission lines; however, blended or doublet emission lines (e.g. \NIIb\ + \mbox{\Ha\ $\lambda$6563} + \NIIa, \mbox{\SII\ $\lambda \lambda$6717, 6731}) complicate the analysis significantly (see \citealt{Schirmer13} for an example of deblending these lines).  

Parametric analysis relies on the assumption that the observed emission line profiles can be reproduced using a linear combination of analytic line profiles (most often Gaussian). Any number of Gaussian components can be fit to each emission line, and each component is parametrized by its velocity (relative to the systemic velocity of the galaxy), velocity dispersion and peak flux. Parametric emission line fitting codes overcome difficulties associated with blended lines and doublets by fitting the profiles of all emission lines simultaneously. Parametric analysis is most suited to systems where multiple kinematic components are superimposed along the line of sight and all contribute significantly to the total line emission of the galaxy. A fitted spectrum is produced for each kinematic component, allowing the physical properties of different line-emitting regions to be analysed independently. 

The kinematic complexity of J2240 (revealed by VLT/XSHOOTER long-slit data; see \citealt{Schirmer13}) suggests that parametric emission line analysis would provide significant insight into the ionizing sources of this system. We use a modified version of the IDL emission-line fitting toolkit \textsc{lzifu} (Ho et al., in prep, see Section 4.1 of \citealt{Ho14} for a brief description) to measure the fluxes, velocities and velocity dispersions for a suite of 14 temperature-, density- and excitation-sensitive emission lines (\mbox{\OII\ $\lambda\lambda$3726, 3729}, \mbox{[Ne III] $\lambda$3869}, \Hg, \OIIIc, \mbox{\HeII\ $\lambda$4686}, \Hb, \OIIIb, \OIIIa, \NIIb, \Ha, \NIIa, \SIIa\ and \SIIb). Unfortunately, the \mbox{\OI\ $\lambda$6300} line lies at a redshifted wavelength of \mbox{8354 \AA} and therefore does not fall within any of our selected spectral channels. \textsc{lzifu} uses the Levenberg-Marquardt least-squares fitting technique (MPFIT; \citealt{Markwardt09}) to model emission line profiles as a superposition of up to three Gaussian components. The reduced-$\chi^2$ values of the 1, 2 and 3 component fits are compared to determine the number of statistically significant components in each spaxel. We require the \Ha\ and \OIII\ fluxes of each significant component to be detected at the 3$\sigma$ level. All emission line maps shown in this paper only include spaxels detected at the 3$\sigma$ level.

The flexibility of \textsc{lzifu} allows it to effectively reproduce a large variety of complex line profiles across J2240, as illustrated in Figure \ref{Fig:schematic}. This schematic shows the total reduced spectra (black) and emission line fits (individual kinematic components shown in red, blue and green) as well as the total fits (brown) for eight different regions in J2240. Red numbers in the top right hand corner of each sub-plot correspond to the numbered locations in the central image of the galaxy. The two upper panels in each sub-plot show spectra zoomed in on the \Hb\ + \OIII\ complex (4800-5100 \AA; left hand side) and the \NII\ + \Ha\ and \SII\ complexes (6450-6800 \AA; right hand side). These spectral regions contain the strongest emission lines in the wavelength range probed by our data and therefore demonstrate the complexity and variety of the line profiles most effectively. The two lower panels show the fitting residuals for each spectral region. The fluxes are normalized to the peaks of the \OIII\ and \Ha\ lines in each spectral range individually. The central panel shows the intensity within the central velocity slice of the \OIIIa\ emission line. 

In referring to different regions of the galaxy, we adopt the nomenclature of \citet{Schirmer13} whereby the south western region of the system is labelled the `cloud' and the north eastern side the `disc'. We also note the presence of two prominent emission peaks in the cloud region of the galaxy, which we label the northern and southern \OIII\ peaks. 

We observe a range of line profiles across the system. We note that the red channel is much noisier than the green channel due to the prominence of skyline residuals at large observed wavelengths (e.g. $\lambda_{H\alpha, obs} \sim 8700 \AA$). Near the outer edge of the cloud (region 3), the emission lines are very narrow and only a single kinematic component is present. The \OIII, \Ha\ and \NIIa\ emission lines have large EWs (247, 47 and 18 \AA\ respectively), but \SII, \NIIb\  and \Hb\ are barely detected above the noise. The remainder of the probed regions lie closer to the main body of the galaxy and therefore the \NII, \Ha, \OIII, \Hb\ and \SII\ lines all have large EWs (ranging from tens to hundreds of Angstroms). The emission line fitting indicates that the spectra of these regions are best fit by a superposition of two narrow components with a small (\mbox{$\sim$ 50 $\rm km \, s^{-1}$}) velocity separation, plus a prominent broader component whose velocity relative to the narrow components varies between different regions of the galaxy. 

The contribution of the stellar continuum emission to the total observed flux is negligible across the majority of the system ($< 3$ per cent of the peak \Ha\ flux and $< 0.7$ per cent of the peak \OIII\ flux). We detect significant stellar continuum emission (7.7 per cent of the peak \Ha\ flux and 1.9 per cent of the peak \OIII\ flux) in region 6, indicating that this is a region of high stellar density compared to the majority of the system (see Section \ref{subsec:continuum} for further discussion). 

\begin{figure}
\centerline{\includegraphics[scale=0.5]{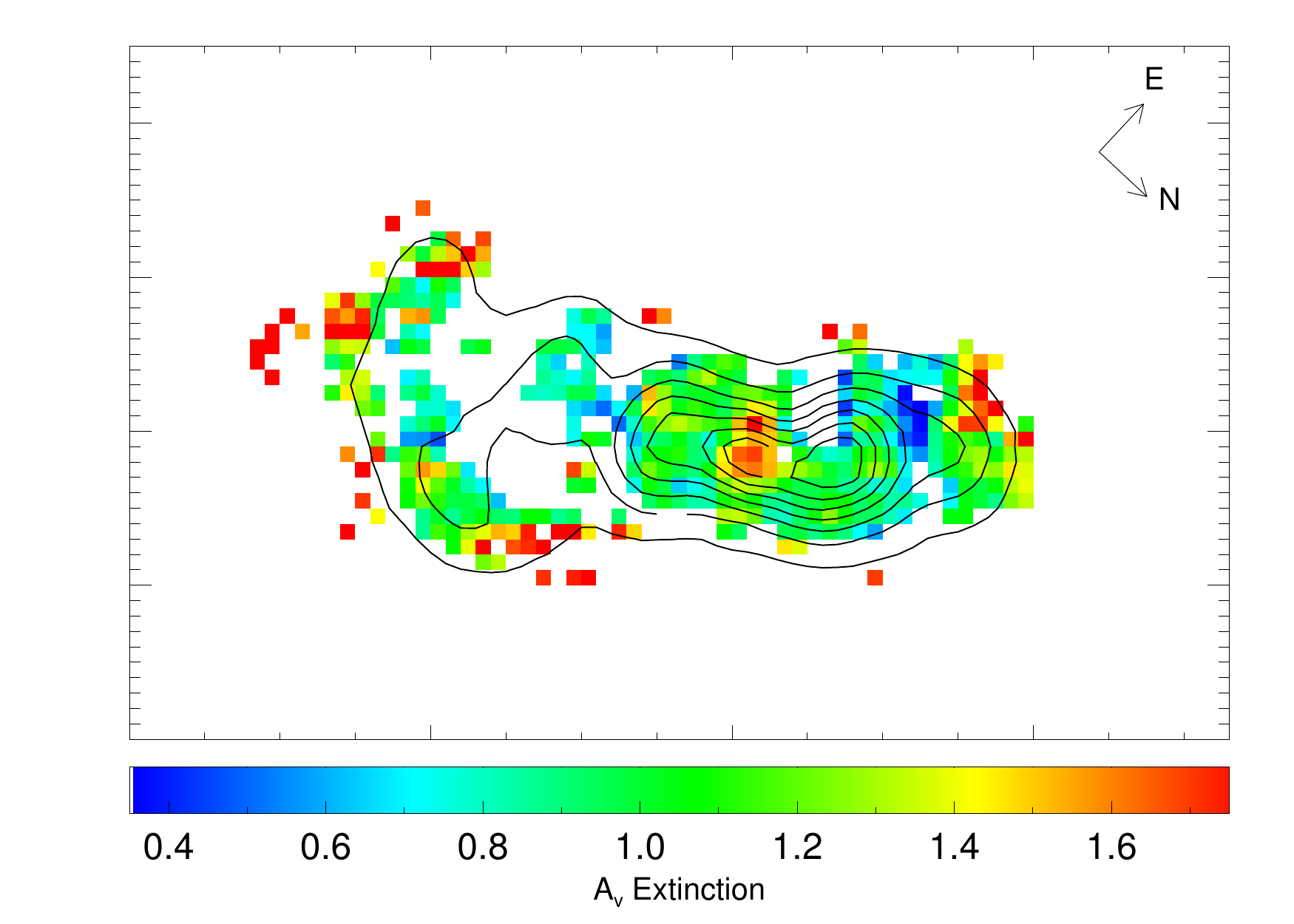}}
\caption{Map of the V-band extinction ($A_v$) along different lines of sight to J2240, derived using the Balmer decrement and assuming an unreddened \Ha/\Hb\ ratio of 2.86. Contours trace the total \OIII\ emission (summed across all kinematic components). The majority of the \OIII-emitting region of the galaxy has \mbox{$A_v <$ 1.25 mag}, with the exception of the nucleus which is more heavily attenuated (\mbox{$A_v \sim$ 1.75 mag}). 
}\label{Fig:extinction}
\end{figure}

It is clear that the emission line spectra of J2240 are well characterised by the superposition of multiple Gaussian components. The fitting residuals under the \OIII\ lines (\mbox{\NII\ + \Ha} complex) are less than 1 (7) per cent of the peak \OIII\ (\Ha) flux. The difference in the size of the residuals between spectral regions is not indicative of poorer quality fits to the \NII\ + \Ha\ complex, but reflects the increase in continuum level fluctuation with wavelength. The residuals under the \mbox{\NII\ + \Ha} complex are less than 2.5 times the standard deviation of the underlying continuum, whereas this number can be as large as 9 under the \OIII\ lines due to their large equivalent widths.

The quality of the fits to the \SII\ doublet indicates that the skylines do not compromise the accuracy of our emission line fitting in the red channel. However, the derived errors on the \SII\ line fluxes are unrealistically high due to the presence of a skyline in the middle of the doublet. We therefore accept or reject the fits to the \SII\ line based on the signal-to-noise of the \NIIa\ line. 

\begin{figure*}
\centerline{\includegraphics[scale=1, clip = true, trim = 0 110 0 95]{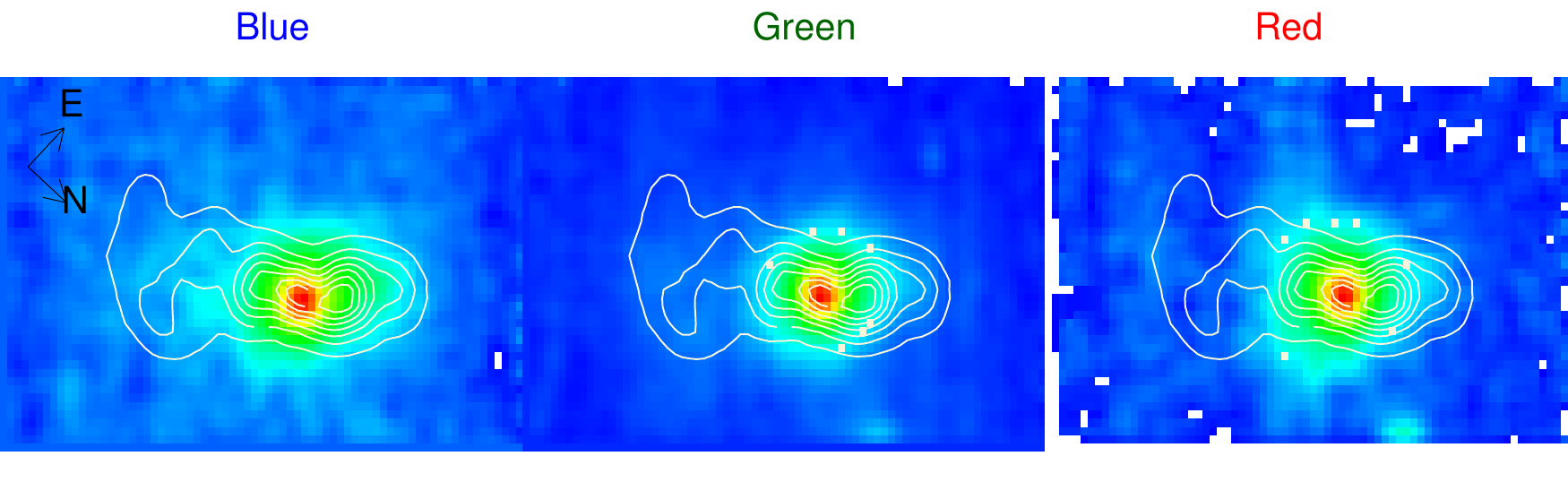}}
\caption{Maps of the integrated continuum emission in the (left to right) blue, green and red channels. Contours trace the total \OIII\ emission. The continuum reveals the underlying stellar body of the galaxy, which is mostly rounded in shape but somewhat elongated towards the north eastern (right) side of the system. There is very little continuum emission in the vicinity of the bright line-emitting cloud, suggesting that this ionized gas cloud may be spatially distinct from the main body of the galaxy.}\label{Fig:continuum}
\end{figure*}

\begin{figure*}
\centerline{\includegraphics[scale=1,clip=true,trim=0 235 0 0]{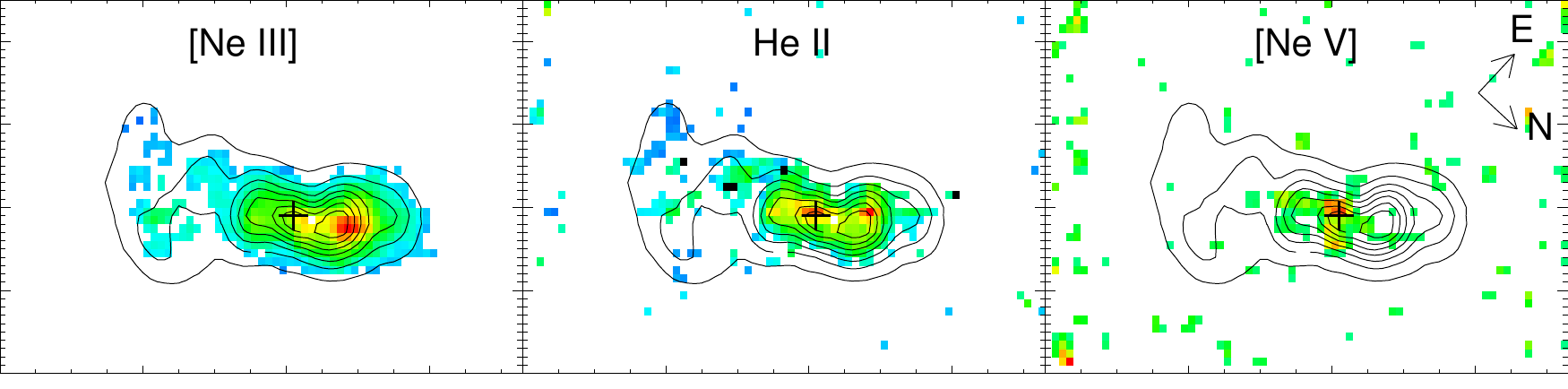}}
\caption{Maps of \NeIII\ $\lambda$3869, \HeII\ $\lambda$4686 and \NeV\ $\lambda$3426 flux. Contours trace the total \OIII\ emission. Black plus signs mark the location of the continuum peak. The \NeIII\ flux has very similar spatial distribution to \OIII, which is expected given that the difference in their required ionization energies is only \mbox{5.84 eV}. There are two peaks in the \HeII\ emission: one coincident with the \NeIII\ peak and one coincident with the continuum peak. The \NeV\ emission, requiring the highest ionization energy, is spatially confined to a small region which allows us to pinpoint the location of the central AGN. Other regions with prominent \HeII\ and/or \NeIII\ emission are likely to be ionized by shocks.}\label{Fig:AGN_lines}
\end{figure*}

\subsection{Extinction correction}
\label{subsec:extcor}
We calculate the V-band extinction ($A_V$, in magnitudes) along the line of sight to each spaxel using the \Ha/\Hb\ flux ratio. The hydrogen Balmer series originates from recombination to the \mbox{n = 2} level of hydrogen. The unreddened intensity ratios between different lines in the series are determined by their transition probabilities which vary depending on the temperature and density of the line-emitting gas as well as the shape of the ionizing radiation field. The extinction along any line of sight can be calculated by comparing measured Balmer line ratios with their intrinsic values.  

We convert the \Ha/\Hb\ ratio in each spaxel to an $A_V$ value using the \citet{Fischera05} extinction curve with $R_V^A = 4.5$ (see \citealt{Vogt13} for a summary of the calculation method). We adopt an unreddened \Ha/\Hb\ ratio of 2.86, appropriate for Case B recombination in a nebula with an electron temperature of \mbox{10,000 - 20,000 K} and an electron density of \mbox{$\sim$100 $\rm cm^{-2}$} \citep{Osterbrock06}. The electron temperature and density ranges are selected based on the findings of \citet{Schirmer13}.

The resulting $A_V$ map is shown in Figure \ref{Fig:extinction}. Contours in this and all subsequent figures trace the total \OIII\ emission (summed across all kinematic components) unless otherwise stated. The majority of the \OIII-emitting region of the galaxy has \mbox{0.75 $< A_v <$ 1.25 mag}, with the exception of the nucleus which is more heavily attenuated (\mbox{$A_v \sim$ 1.75 mag}). The disc region is more heavily attenuated than the cloud, consistent with the findings of \citet{Schirmer13}. 

We also calculate extinction values using an increased unreddened Balmer decrement of 3.1 which accounts for the collisional excitation of \Ha\ in the presence of the AGN ionizing radiation field \citep{Gaskell84}. This systematically decreases the calculated extinction values by \mbox{0.3 mag}. Throughout this paper we correct line fluxes for extinction using both sets of extinction values and discuss the impact on our derived quantities.

\section{Ionization structure and sources: building up the picture of J2240}
\label{sec:initial_analysis}
Our emission line fitting indicates that several kinematic components contribute significantly to the emission signature of J2240. However, determining the geometry and ionization source of each kinematic component can be a non-trivial task, especially for systems in which the line profiles and dominant ionizing source(s) vary significantly across the galaxy. In the following section, we use integrated continuum maps to trace the stellar body of the galaxy and map the flux in high ionization energy emission lines to pinpoint the location of the central AGN. We then compare maps of the continuum and \OIIIa\ fluxes over several velocity slices covering the full dynamical range of the system. This allows us to dissect the spatial extent and relative contribution of individual ionizing sources as a function of both wavelength and position in the galaxy. 

\subsection{Continuum properties}
\label{subsec:continuum}
CFHT imaging and VLT/XSHOOTER spectra of J2240 suggest that there is a large amount of ionized gas on the south western side of the system which is spatially distinct from the main body of the galaxy \citep{Schirmer13}. We investigate the location and spatial extent of the stellar component of J2240 by mapping the integrated continuum emission in each of the spectral channels (see Figure \ref{Fig:continuum}). The clear continuum peak reveals the location of the galaxy centre, and is coincident with one of the two \OIII\ peaks evident from the contours. The continuum flux decreases smoothly with distance from the centre of the galaxy. An elongation is visible in the continuum emission towards the north eastern side of the system. A similar feature is seen in the \OIII\ emission, suggesting that the system may have been morphologically disturbed during recent galaxy-scale interactions. We note that there is very little continuum emission on the south western side of the system, despite the strength of the nebular emission lines observed in this region (see e.g. Figure \ref{Fig:alignment}). The limited extent of the stellar continuum emission is consistent with the hypothesis that the ionized gas cloud extends beyond the stellar body of the galaxy.

\begin{figure*}
\centerline{\includegraphics[scale=1.6,clip=true,trim=0 0 250 0]{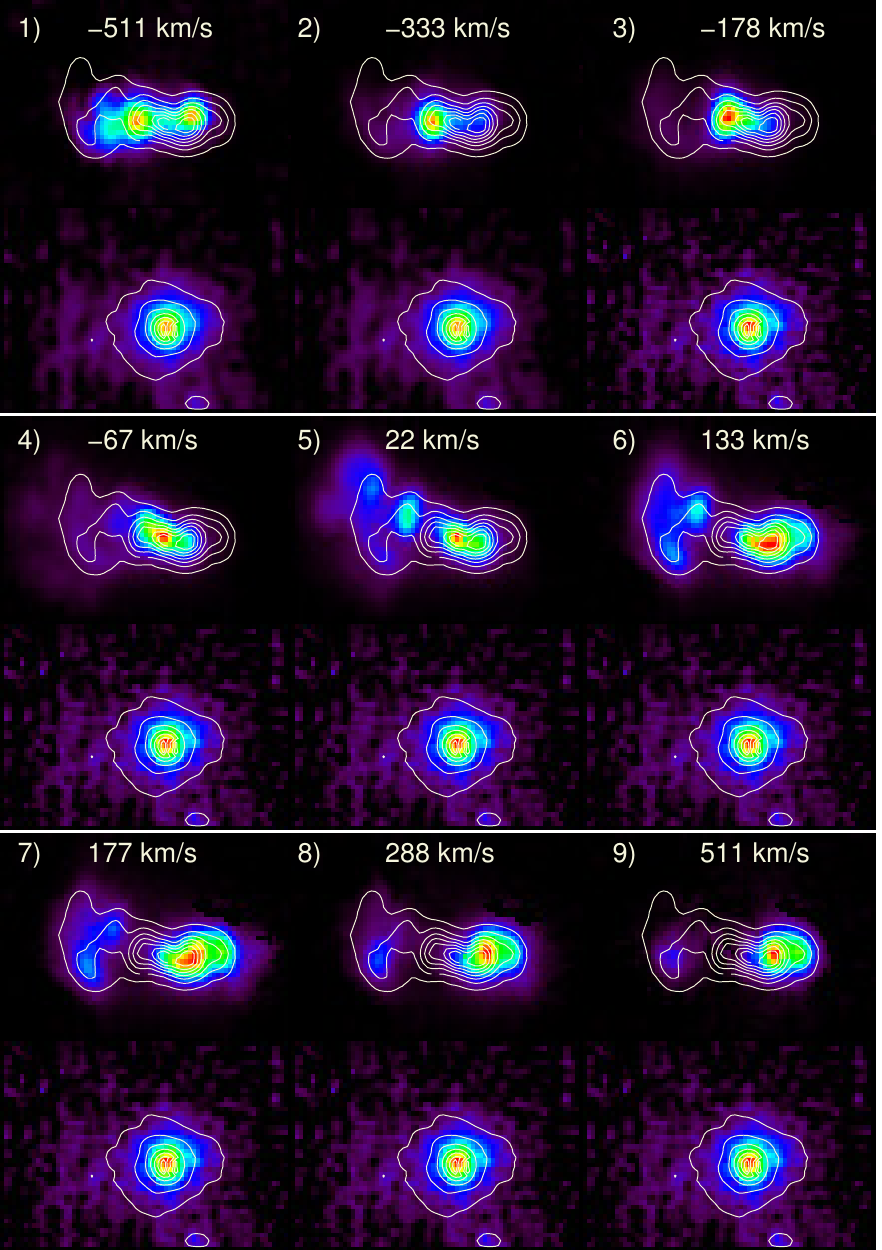}}
\caption{Flux maps for nine velocity channels across the \OIIIa\ emission profile, with continuum maps extracted from the same channels for comparison. Contours on the emission line maps trace the total \OIII\ emission, and contours on the continuum maps trace the integrated continuum emission in the red channel. It is clear that the morphology of the continuum emission does not evolve significantly as a function of wavelength. In contrast, the morphology of the \OIII\ emission evolves dramatically between velocity slices. The highest velocity gas (slices 1, 2, 8 and 9) is primarily aligned along the major axis of the continuum emission suggesting that it may be associated with the galaxy's underlying stellar population. However, the central velocity slices are dominated by line emission from the gas cloud which is misaligned by $\sim$ 45\degree\ compared to the continuum axis.}\label{Fig:channels}
\end{figure*}

\begin{figure*}
\centerline{\includegraphics[scale=1,clip=true,trim=0 120 0 0]{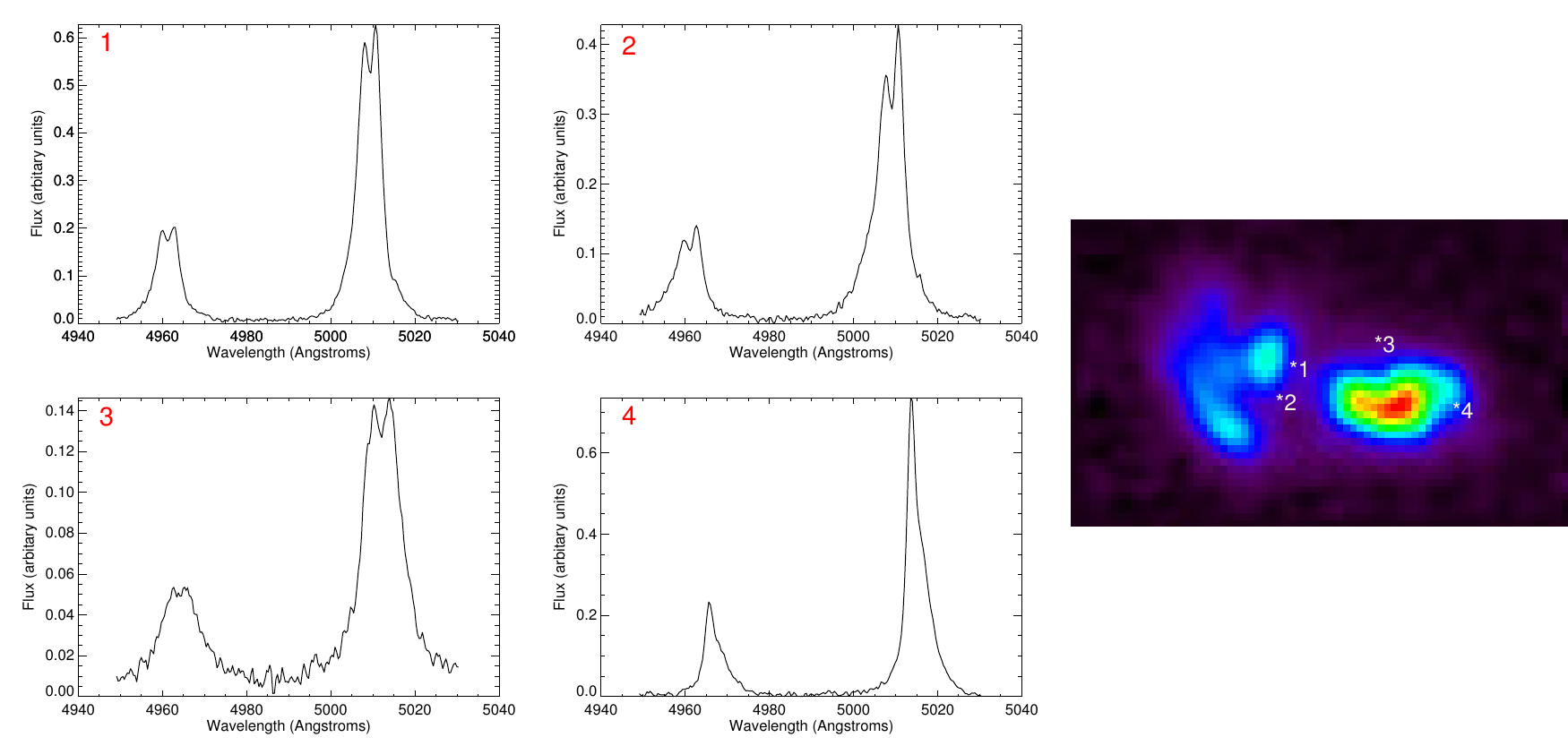}}
\caption{Spectra centred around the \OIIIa\ and \OIIIb\ emission lines, highlighting the superposition of two narrow components along various edges of the galaxy disc. The second component is associated with the galaxy cloud. The velocity offset is due to the differing velocity amplitudes of the cloud and disc rotation curves.}\label{Fig:narrow_comps}
\end{figure*}

\subsection{Uncovering the AGN using \HeII, \NeIII\ and \NeV}
\label{subsec:AGN_lines}
Emission line ratios extracted from the narrow line region of J2240 indicate that the central engine is likely to be powered by an AGN \citep[see e.g. appendix A of][]{Schirmer13}. However, a single 1D VLT/XSHOOTER spectrum across the major axis of the galaxy is insufficient to deduce the location of the central AGN, which could be offset from the slit axis. Our IFU data eliminate this issue by allowing us to investigate how the hardness of the ionizing radiation field varies across the entire optical extent of the galaxy. Doubly ionized oxygen (excited by an ionization energy of \mbox{35.12 eV}) is detected across a very spatially extended region of J2240, but probing species excited by larger ionization energies allows us to pinpoint the hardest ionizing sources in the galaxy. In particular, we investigate the distribution of flux in \NeIII\ $\lambda$3869, \HeII\ $\lambda$4686 and \NeV\ $\lambda$3426, which are excited by ionization energies of \mbox{40.96 eV}, \mbox{54.4 eV} and \mbox{97.12 eV} respectively. Neon is a relatively abundant element, produced by carbon burning in the late stages of the evolution of massive stars \citep{Levesque14}. The strength of the \NeIII\ emission line correlates with the \OIII\ flux, and therefore \NeIII\ can be used in place of \OIII\ in AGN diagnostics \citep{Perez-Montero07}. However, the relatively low ionization energy of [Ne~II] makes \NeIII\ sensitive to slow and fast shocks. \NeV\ requires a very large ionization energy and is therefore only detectable in the immediate vicinity of an AGN or in material with shock velocity greater than \mbox{500 km $\rm s^{-1}$} \citep{Allen08}. Helium is also a very abundant element, produced from hydrogen burning during the main sequence lifetime of stars. \HeII\ can be excited by AGN activity, fast shocks and young, hot stars (such as O-stars or Wolf-Rayet stars). Given the weakness of the stellar continuum emission in J2240, it is unlikely that active star-formation contributes significantly to the \HeII\ emission. \HeII\ is excited by an ionization energy between that of \NeIII\ and \NeV, and therefore we expect the \HeII\ flux distribution to be intermediate between that of the two neon lines.

We show maps of the total \NeIII, \HeII\ and \NeV\ emission in Figure \ref{Fig:AGN_lines}. (It is not necessary to break the flux into components because these emission lines only originate from the most energetic ionizing sources.) The spatial distribution of the \NeIII\ emission is very similar to that of the total \OIII\ emission within the main body of the galaxy, suggesting that the ionization parameter is high enough across the galaxy to excite both lines equally. However, the \NeIII\ emission in the cloud is concentrated around the northern and southern \OIII\ peaks, suggesting that these regions may be shock-excited. 

The \HeII\ flux is more concentrated towards the centre of the galaxy than the \NeIII\ flux, and has a double-peaked distribution. One of the peaks is coincident with the \NeIII\ peak, whilst  the second peak is coincident with the continuum peak. We also see evidence for \HeII\ emission in the southern \OIII\ peak of the cloud, strengthening the case for shock excitation in this region. The \NeV\ emission is very spatially concentrated around the continuum peak. The presence of a very hard ionizing radiation field at the centre of the galaxy is consistent with AGN activity. The second \HeII\ peak is invisible in \NeV, and is therefore likely to be powered  by fast shocks. There is no evidence for a second active galactic nucleus from this data. 

\subsection{\OIIIa\ velocity channel maps}
\label{subsec:channel_maps}
We probe the spatial and spectral variation in the emission line morphology of J2240 by constructing flux maps for nine velocity channels across the \OIIIa\ emission line profile (see Figure \ref{Fig:channels}). Maps of the continuum emission across the same channels have been included below each emission line map for comparison. No signal-to-noise ratio cuts have been applied. Contours on the emission line maps trace the total \OIII\ emission, whilst contours on the continuum maps trace the integrated continuum emission from the red channel. The colorbar is scaled independently in each map and velocity slice to span the full range of flux levels, from zero flux in black to maximum flux in red.

It is clear that the morphology of the continuum emission does not evolve significantly as a function of wavelength. This reflects the fact that the continuum emission can be explained by a single, kinematically simple emission source - the underlying stellar population of the galaxy. In contrast, the morphology of the \OIII\ emission evolves dramatically between velocity slices. The highest velocity gas (slices 1, 2, 8 and 9) is primarily aligned along the major axis of the continuum emission, suggesting that it may be associated with the underlying stellar population of the galaxy (i.e. tracing the gas disc of the galaxy). However, the central velocity slices are dominated by line emission from the gas cloud which is misaligned by $\sim$ 45\degree\ compared to the continuum axis. The emission from the cloud is clumpy and morphologically irregular, suggesting that there may be significant variation in the ionization parameter and/or gas density within the cloud.

The evolution in emission morphology as a function of wavelength suggests that a) the rotation curve of the galaxy disc has a larger velocity amplitude than the rotation curve of the cloud structure, and b) the cloud component has a much larger \OIII\ luminosity than the south western-most region of the disc, causing the former to be more prominent in the central velocity slices. Slices four and five have the largest \OIII\ fluxes, and both are dominated by \OIII\ emission from the continuum peak, consistent with our hypothesis that this is the location of the central AGN. 

The channel maps reveal several kinematically broad regions which emit strongly across several velocity channels. The elongation on the north eastern side of the galaxy is seen in the last four velocity slices, the emission line peak between the disc and the cloud is seen prominently in the first three velocity slices, and the northern \OIII\ peak in the cloud is observed in the last five velocity slices. This suggests that there may be several localised regions of energetic, high temperature gas in J2240. 

Our analysis suggests that at least three ionizing sources contribute significantly to the line emission of J2240 - the gas disc, which extends along the major axis of the galaxy's continuum emission, the gas cloud, which is ionized along a distinctly misaligned axis, and more energetic ionizing source(s) which increase the turbulent energy of the gas. This is consistent with the line profiles and emission line fits presented in Figure \ref{Fig:schematic}. We expect the two narrow components to be superimposed at the edge of the gas disc, and observe double-peaked line profiles in several locations (see Figure \ref{Fig:narrow_comps}). The two narrow components are offset in velocity due to the different velocity amplitudes of the disc and cloud rotation curves, but appear to have very similar velocity dispersions. Double peaked line profiles are not visible in the centre of the galaxy because emission from the galaxy component is much stronger than emission from cloud component in this region. 

\subsection{Component assignment}
\begin{figure}
\centerline{\includegraphics[scale=0.45]{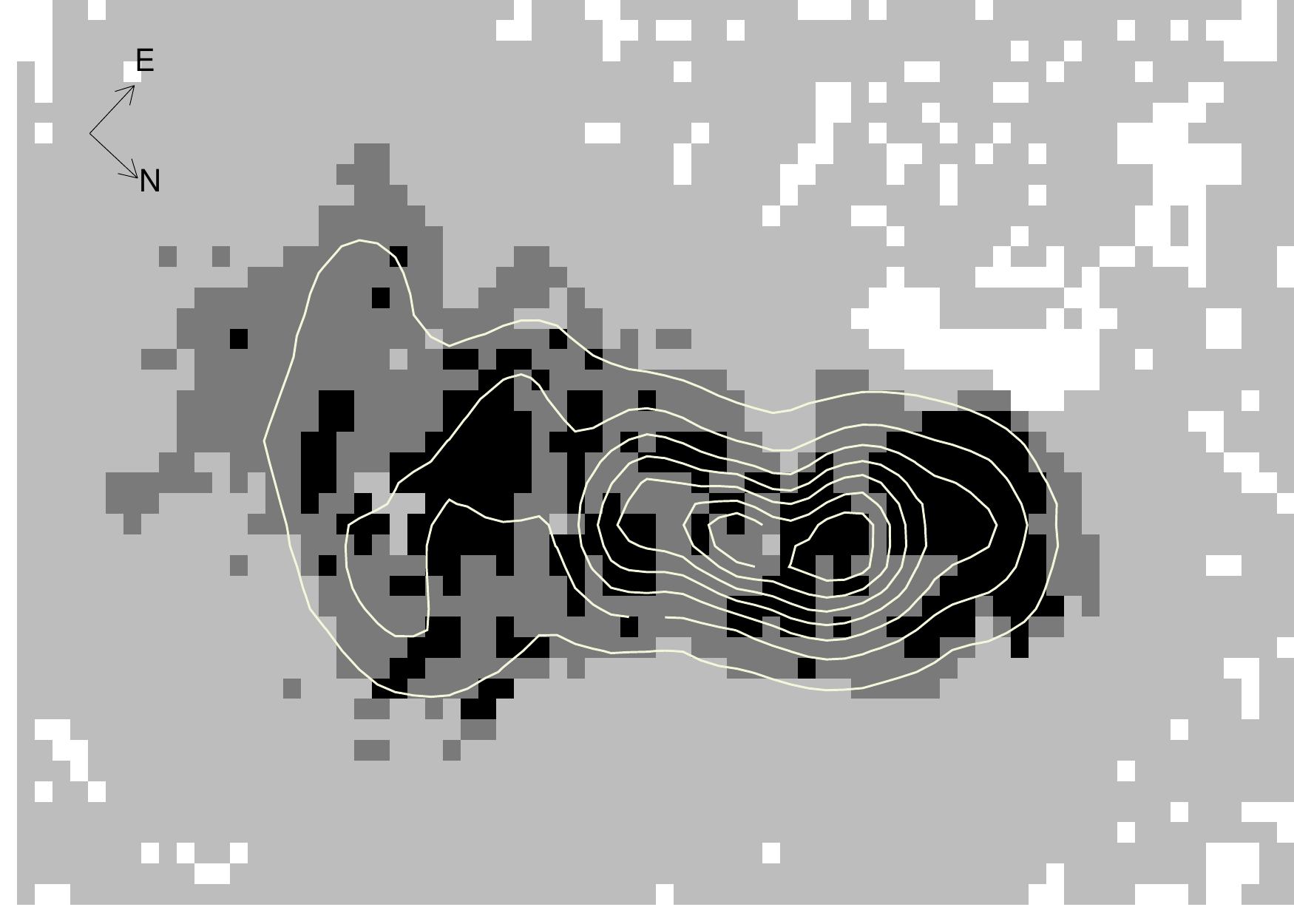}}
\caption{Map showing the number of statistically significant kinematic components in each spaxel of J2240, as determined by comparing the reduced $\chi^2$ values of the 1, 2 and 3 component fits. Spaxels requiring 1, 2 and 3 components are coloured light grey, dark grey and black respectively. Spaxels for which the emission line fitting did not converge are not coloured. Contours trace the total \OIII\ emission. At least two kinematic components are required to reproduce the observed spectra across the majority of the \OIII-emitting region, and several regions which are likely to be shock-excited require three components.}\label{Fig:component_map}
\end{figure}

Figure \ref{Fig:component_map} shows the number of statistically significant kinematic components in each spaxel of J2240 (determined as described in Section \ref{subsubsec:schematic}). Spaxels requiring one, two and three components are coloured light grey, dark grey and black respectively. Spaxels for which the emission line fitting did not converge are not coloured. At least two kinematic components are required to reproduce the observed spectra across the majority of the \OIII-emitting region, as expected from our analysis of the \OIII\ channel maps. Three kinematic components are required in the vicinity of the elongation, at the \NeIII\ peak, in the region between the cloud and the main body of the galaxy and at the northern \OIII\ peak of the cloud. Our analysis of the channel maps has revealed that 3/4 of these regions have large velocity dispersions, and the \NeIII\ peak is detected in \NeIII\ and \HeII, suggesting that energy injection from shocks is likely to drive the gas turbulence in all four regions.

The subsequent sections of this paper will delve into the physical properties of each of the individual kinematic components, making it imperative that the components are numbered consistently across the galaxy according to their physical origin. \textsc{lzifu} automatically numbers the components in order of increasing velocity dispersion within each individual spaxel. This guarantees that the broad components identified in the channel maps will always appear in component 3. The two narrow components have similar velocity dispersions, but the velocity dispersion in the disc is consistently larger than that in the cloud, ensuring that component 1 will be dominated by emission from the ionized cloud and component 2 will trace the rotation curve of the main body of the galaxy as well as some of the turbulent gas in the cloud. 

\section{3D spectroscopic analysis}
\label{sec:3D_analysis}

\subsection{\OIII\ morphology}
\begin{figure*}
\centerline{\includegraphics[scale=1,clip=true,trim=0 240 0 0]{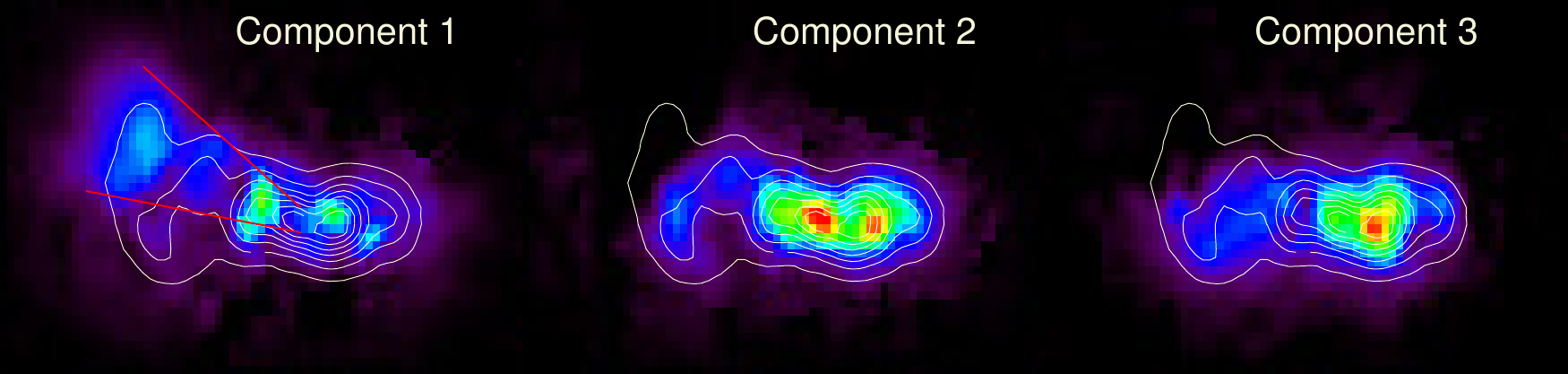}}
\caption{\OIII\ flux maps for each of the three line-emitting components. Contours trace the total \OIII\ emission. The kinematic components have different \OIII\ morphologies, suggesting that they have different physical origins. Component 1 traces emission from the ionized gas cloud, component 2 traces the rotation curve of the main body of the galaxy and component 3 traces turbulent gas energized by fast shocks. Red lines overplotted on the component 1 map trace the edges of the possible AGN ionization cone.}\label{Fig:OIII_morphology}
\end{figure*}

Figure \ref{Fig:OIII_morphology} shows \OIII\ flux maps for each of the individual kinematic components. The flux range spanned by the colorbar is fixed across all three maps to indicate the relative contribution of different kinematic components and physical regions to the overall \OIII\ emission of the system. Component 1 clearly traces emission from the ionized gas cloud, as expected. There is a large amount of diffuse gas with relatively weak \OIII\ emission, and the regions of most intense \OIII\ flux appear to be strongly associated with the misaligned emission axis identified from the channel maps. This emission axis intersects with the main body of the galaxy at the location of the AGN, suggesting that we may be observing an AGN ionization cone (indicated by the red lines in Figure \ref{Fig:OIII_morphology}).

The majority of the \OIII\ emission in component 2 originates from the main body of the galaxy. The spatial distribution of \OIII\ flux in this component is very similar to the total \OIII\ flux distribution traced by the contours, indicating that the second component is the dominant source of \OIII\ flux across the majority of J2240. The \OIII\ emission peaks strongly at the location of the central AGN. We also clearly recover the northern and southern \OIII\ peaks in the cloud, which have both been identified as possible shocked regions. 

The \OIII\ emission in component 3 originates from a similar region of the system to component 2, but with significantly different morphology. The \NeIII\ peak is the dominant source of \OIII\ emission in component 3, and has a similar peak flux to the AGN emission in component 2. The strong \OIII\ emission from the \NeIII\ peak appears to drive the strong warp in the \OIII\ contours. The elongation on the north eastern side of the galaxy is clearly revealed in this component, and appears to be physically distinct from the remainder of the galaxy. We also observe a region of weaker, diffuse emission between the north western edge of the galaxy disc and the northern \OIII\ peak of the cloud. This feature is not observed in either the channel maps or the high ionization line maps, and is only revealed when the much stronger narrow components are removed. 

\begin{figure*}
\centerline{\includegraphics[scale=1,clip=true,trim=0 140 0 0]{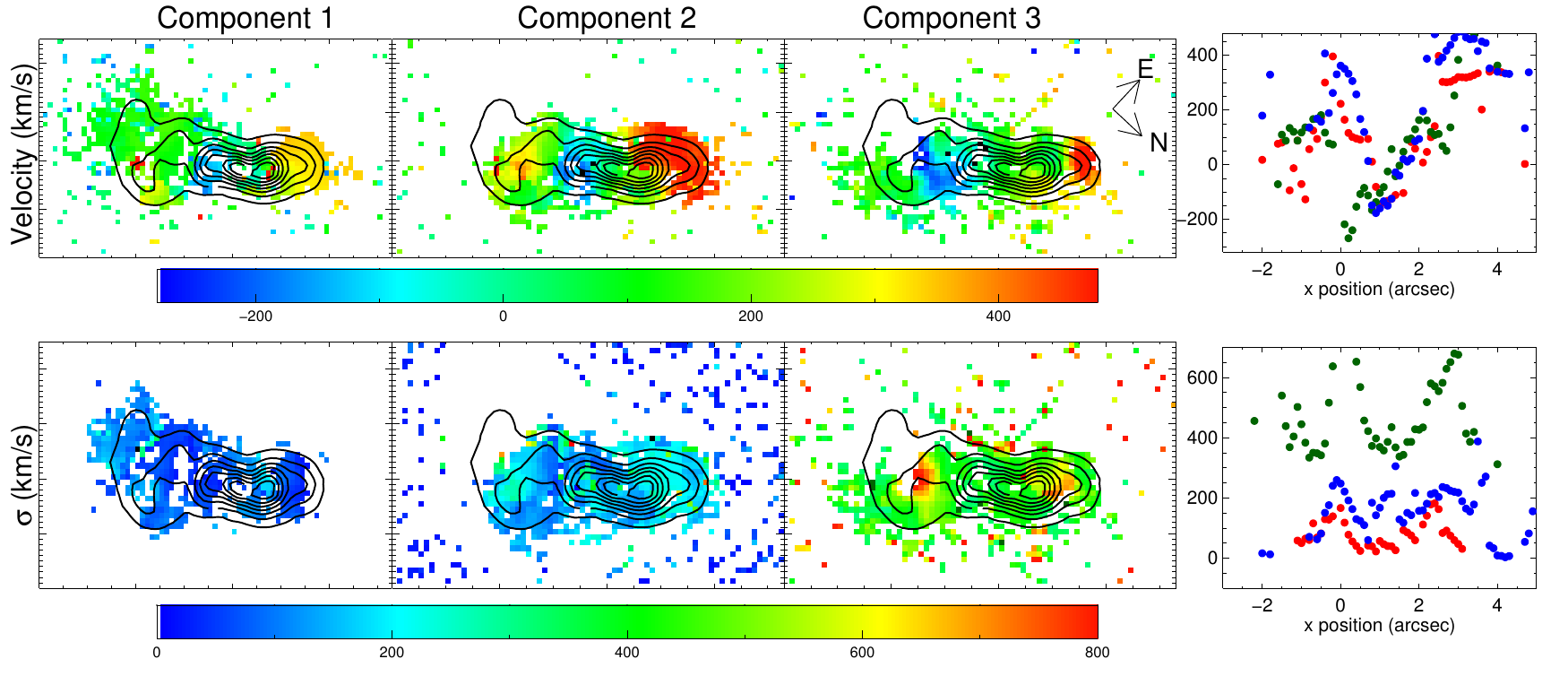}}
\caption{Maps of the velocity field and velocity dispersion ($\sigma$) for each of the three line-emitting components, and position-velocity and position-$\sigma$ diagrams extracted along the major axis of the galaxy. Contours trace the total \OIII\ emission. In the two right hand panels, positions are with respect to the central AGN and components 1, 2 and 3 are shown in red, blue and green respectively. All three components show clear rotation signatures, but the gas is globally disturbed. The velocity dispersion reaches \mbox{600-800 $\rm km \, s^{-1}$} in some regions of component 3, indicating energy injection from fast shocks and outflowing material.}\label{Fig:vel_maps}
\end{figure*}		

\begin{figure*}
\centerline{\includegraphics[scale=0.95]{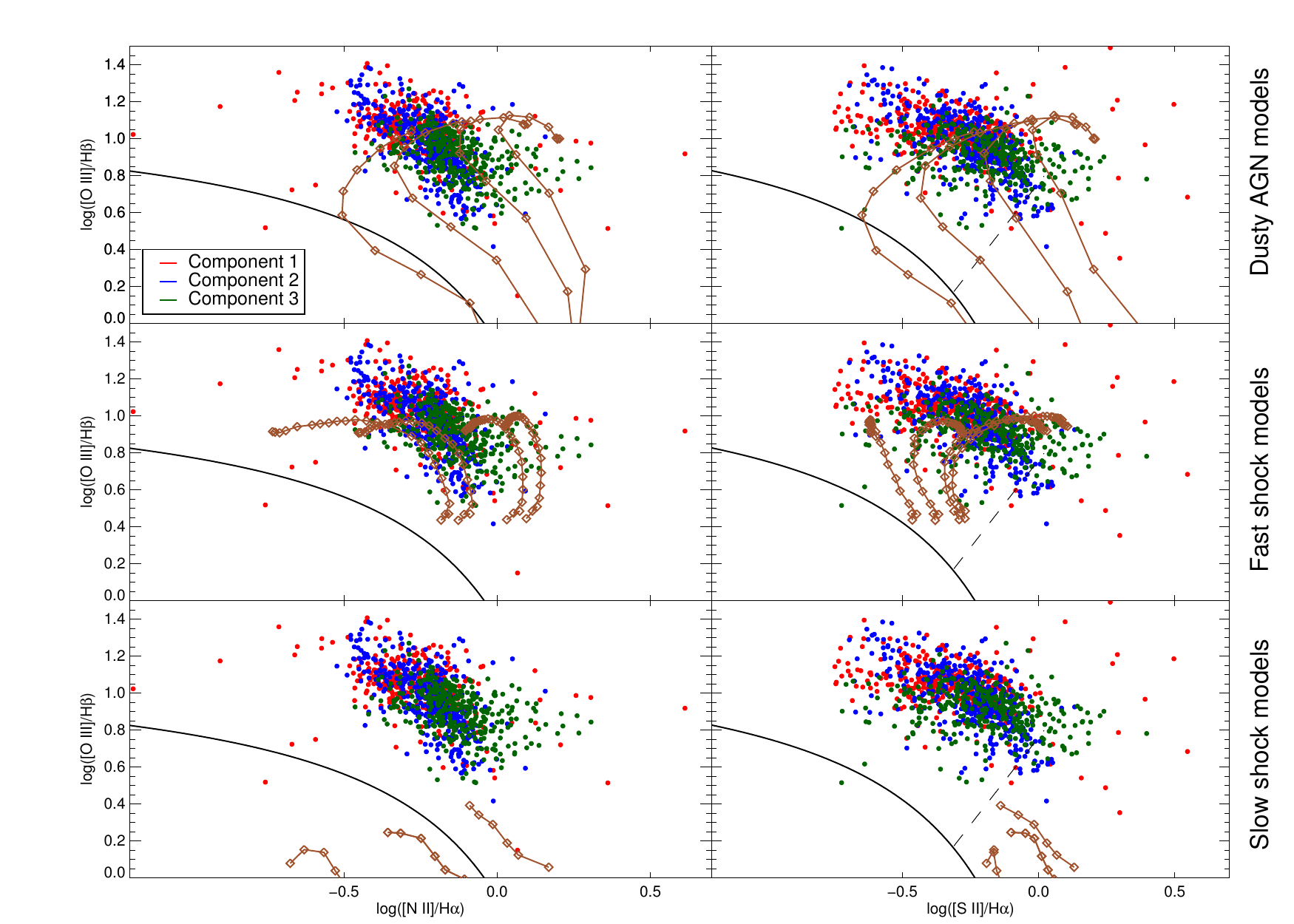}}
\caption{\NIIHa\ and \SIIHa\ vs. \OIIIHb\ diagnostic diagrams with line ratios extracted for each component of each spaxel. Red, blue and green points are extracted from components 1, 2 and 3 respectively, and brown points have been extracted from model grids. The top panel shows Dusty AGN models \citep{Groves04} with \mbox{12 + log(O/H) = 8.99} and \mbox{$n_e = 100 \, \rm cm^{-3}$}. Solid lines connect points of constant power law index, ranging from \mbox{$\Gamma$ = 1.2 - 2}. The middle panel shows fast shock models \citep{Allen08} with \mbox{12 + log(O/H) = 8.69} and precursor density of \mbox{$n_H = 100 \rm \, cm^{-3}$}. Solid lines connect points of constant magnetic field strength, going from \mbox{10$^{-4}$-100 $\mu$G}. The bottom panel shows slow shock models \citep{Rich10}. Solid lines connect points of consant shock velocity, ranging from \mbox{100-200 $\rm km \, s^{-1}$}. The Dusty AGN models indicate that the anti-correlation between the \NIIHa\ and \OIIIHb\ ratios can be explained by ionization parameter variaions in the line-emitting gas. The fast shock models also coincide with a significant fraction of the data points, indicating that fast shocks could be the principal ionizing source in some regions of the galaxy.}\label{Fig:diagnostic_diagrams}
\end{figure*}

\subsection{Velocity field and velocity dispersion}
Figure \ref{Fig:vel_maps} shows the velocity and velocity dispersion ($\sigma$) fields of each of the kinematic components. We also plot position-velocity and position-$\sigma$ diagrams extracted along the major axis of the galaxy down the right hand column of the figure. Positions are calculated with respect to the central AGN and components 1, 2 and 3 shown in red, blue and green respectively.

The velocity field of component 2 reveals a clear rotation curve over the disc region of the galaxy with an amplitude of \mbox{$\sim$300 $\rm km \, s^{-1}$}. The velocity field of component 1 is also dominated by rotation, but the position-velocity diagram reveals that its rotational velocity is smaller than that of component 2, as expected from our analysis of the channel maps. The co-rotation of component 1 and component 2 suggests that the cloud material  which is superimposed over the main body is enveloping it and is therefore morphologically linked to the main body of the galaxy (as discussed by \citealt{Schirmer13}). The remainder of the cloud material generally lies at the systemic velocity of the galaxy, although some shocked regions in component 2 have more turbulent emission components with anomalous velocities. The emission from the northern \OIII\ peak in the cloud is redshifted by about 200 $\rm km \, s^{-1}$, and there is some evidence for blueshifted material in the north eastern region of the cloud. The average velocity dispersions of components 1 and 2 are \mbox{100 $\rm km \, s^{-1}$} and \mbox{150 $\rm km \, s^{-1}$} respectively, indicating that the gas is globally disturbed despite the ordered rotation we observe.

Component 3 traces the most turbulent gas in the system. The average velocity dispersion is 300-400 $\rm km \, s^{-1}$, but two clumps of gas (in the vicinity of the elongation and between the two \OIII\ peaks in the cloud) have anomalously high velocity dispersions of \mbox{600-800 $\rm km \, s^{-1}$}. These distinct clumps of high velocity dispersion gas are evidence for localised energetic phenomena such as shocks or outflows, whilst the majority of the underlying broad component is likely to be energized by the central AGN. There is some evidence for rotation in component 3; however, the kinematics to the north east of the central AGN are very irregular. A strongly blue-shifted clump of gas is visible just to the south of the \NeIII\ peak, and gas in the vicinity of the elongation appears strongly redshifted.

\begin{figure*}
\centerline{\includegraphics[scale=1,clip = true, trim = 0 70 0 0]{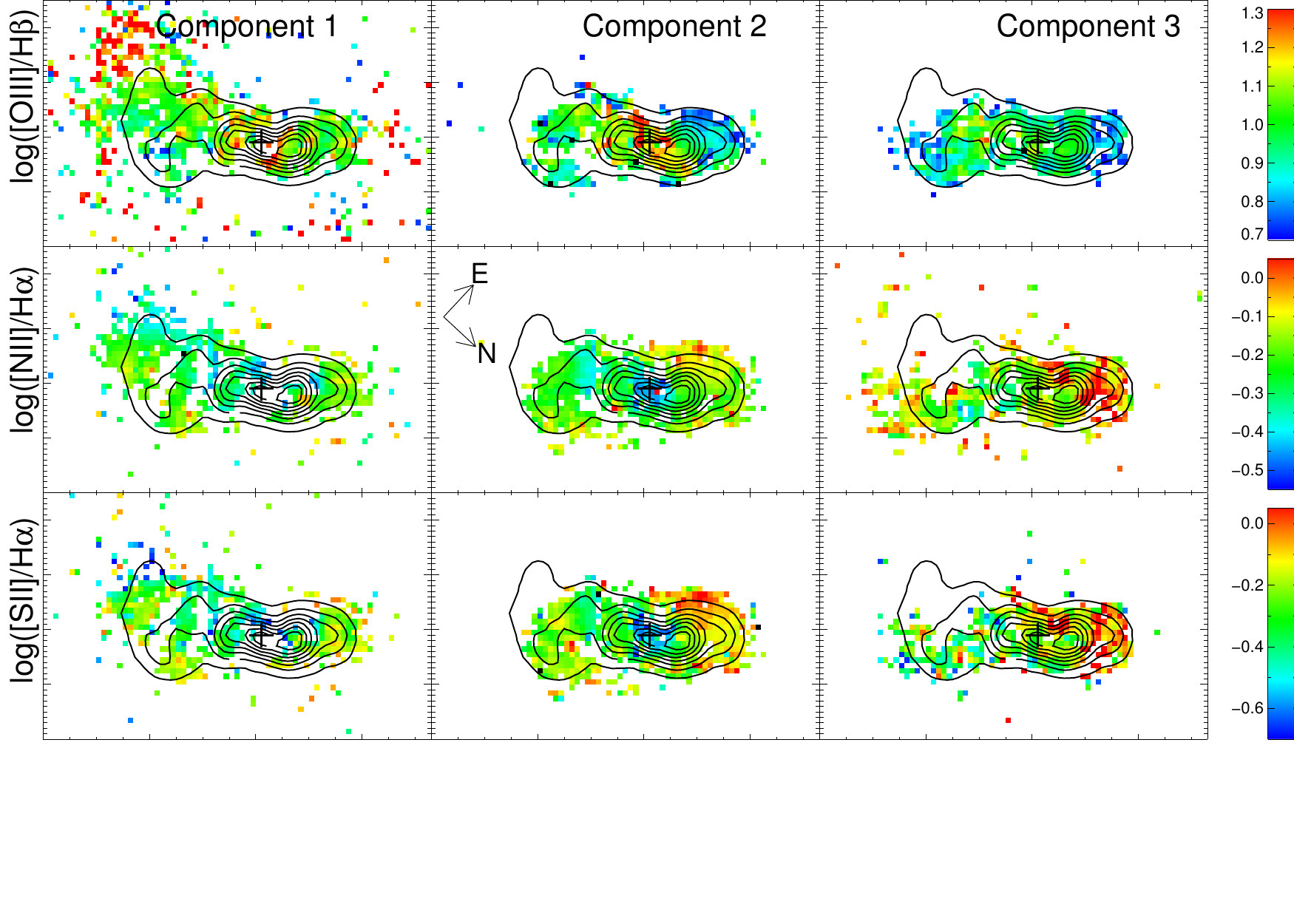}}
\caption{Maps of the \NIIHa, \SIIHa\ and \OIIIHb\ ratios for each of the individual components. Contours trace the total \OIII\ emission. Black plus signs indicate the location of the central AGN. The average \NIIHa\ and \SIIHa\ ratios increase from component 1 to component 3, whilst the average \OIIIHb\ ratio decreases. The largest \NIIHa\ and \SIIHa\ ratios are found on the north eastern side of the system, consistent with our hypothesis that fast shocks are prominent in this region.}\label{Fig:diagnostic_maps}
\end{figure*}

\subsection{Diagnostic diagrams}
\label{subsec:diagnostic_diagrams}
The \NIIHa\ and \SIIHa\ vs. \OIIIHb\ diagnostic diagrams are commonly used to separate star formation from shock excitation and AGN activity \citep{Baldwin81,Veilleux87}. The diagnostic line ratios are virtually insensitive to reddening due to the small wavelength separation of the line pairs (\NIIa\ \& \mbox{\Ha\ $\lambda$6563}, \mbox{\SII\ $\lambda \lambda$ 6717,6731} \& \mbox{\Ha\ $\lambda$6563}, \OIIIa\ \& \mbox{\Hb\ $\lambda$4861}). The \OIIIHb\ ratio is sensitive to the electron temperature and ionization parameter of the line-emitting gas, and the \NIIHa\ ratio is a tracer of metallicity due to the secondary nucleosynthetic production of nitrogen at high oxygen abundances \citep[see e.g.][]{Ke13}. 

Figure \ref{Fig:diagnostic_diagrams} shows \NIIHa\ (left-hand column) and \SIIHa\ (right hand column) vs. \OIIIHb\ diagnostic diagrams with line ratios extracted from the spectrum of each kinematic component in each spaxel of J2240. Components 1, 2 and 3 are shown in red, blue and green respectively. The \citet{Ke01} theoretical upper bound to pure star-formation (solid lines) separates star-formation (bottom left of diagrams with lower \NIIHa, \SIIHa\ and \OIIIHb\ ratios) from harder ionizing sources (top right of diagrams with higher \NIIHa, \SIIHa\ and \OIIIHb\ ratios). All of the observed line ratios for each of the kinematic components lie strongly in the AGN region of the diagnostic diagrams. 

We observe that the \OIIIHb\ ratios are inversely correlated with both the \NIIHa\ and \SIIHa\ ratios. We explore possible origins of this ionization signature by plotting Dusty AGN (top panel; \citealt{Groves04}), fast shock (middle panel; \citealt{Allen08}) and slow shock (bottom panel; \citealt{Rich10}) models over the data, shown in brown. The selected AGN models are calculated for a metallicity of \mbox{12 + log(O/H) = 8.99} (within the 0.9-2$Z_{\odot}$ range derived by \citealt{Schirmer13}) and an electron density of \mbox{$n_e = 100 \rm \, cm^{-3}$} (between the \mbox{$<$50 $\rm cm^{-3}$} density of the cloud and the \mbox{150-650 $\rm cm^{-3}$} density of the disc in J2240; \citealt{Schirmer13}). Solid lines connect points of constant power-law index (increasing from \mbox{$\Gamma$ = 1.2 - 2} as the \NIIHa\ ratio decreases), and the ionization parameter increases from \mbox{$\log U$ = -4} to \mbox{$\log U$ = 0} as the \OIIIHb\ ratios increase along each line. It is clear that ionization parameter variations are able to produce the observed anti-correlation in the line ratios (see Section \ref{subsec:ionization_param} for further discussion of the ionization parameter). Variations in the metallicity may account for the spread in \NIIHa\ ratios at a given \OIIIHb\ ratio. 

The fast shock models are calculated for a metallicity of \mbox{12 + log(O/H) = 8.69} and a pre-shock hydrogen density of \mbox{$n_H = 100 \rm \, cm^{-3}$} (consistent with the metallicity and density ranges derived by \citealt{Schirmer13}). Solid lines connect points of constant magnetic field strength (ranging from \mbox{10$^{-4}$-100 $\mu$G}). The \OIIIHb\ ratios increase as the shock velocity increases from \mbox{300-650 $\rm km \, s^{-1}$} and then decrease again slightly as the shock velocity approaches \mbox{1000 $\rm km \, s^{-1}$}. The fast shock models lie in the same region of the diagnostic diagram as data points with \mbox{log(\OIIIHb) $<$ 1}. This suggests that fast shocks could contribute significantly to the ionization signature in some regions of J2240, consistent with our identification of several prominent shocked regions throughout the system.

The slow shock models are calculated for a range of metallicities (\mbox{8.39 \textless\ 12 + log(O/H) \textless\ 9.29}) and shock velocities (\mbox{100-200 $\rm km \, s^{-1}$}). Solid lines connect points of constant shock velocity, increasing towards larger \NIIHa\ ratios, and the metallicity increases towards larger \NIIHa\ ratios along each line. It is clear that slow shocks are not powerful enough to produce the large \OIIIHb\ ratios we observe, indicating that even if they are present in J2240 they are not the dominant ionizing source.  

\subsection{Maps of diagnostic line ratios}
\label{subsec:diagnostic_maps}
Mapping diagnostic line ratios across galaxy systems reveals spatial variations in the ionization state of the gas. Figure \ref{Fig:diagnostic_maps} shows maps of the \NIIHa, \SIIHa\ and \OIIIHb\ line ratios for each of the kinematic components. The average \NIIHa\ ratios increase from component 1 to component 3 (\mbox{-0.236}, \mbox{-0.211}, \mbox{-0.120}), and the same behaviour is seen in the \SIIHa\ ratios (\mbox{-0.271}, \mbox{-0.234}, \mbox{-0.217}). The average \OIIIHb\ ratios decrease from component 1 to component 3 (1.003, 0.955, 0.882), demonstrating the observed anti-correlation between the \OIIIHb\ ratios and \NIIHa\ and \SIIHa\ ratios. 

The component 1 \OIIIHb\ map has a similar morphology to the \OIII\ flux map in Figure \ref{Fig:OIII_morphology}. There is very little variation in the \OIIIHb\ ratio over the \OIII-bright region of the galaxy (\mbox{0.9 $<$ log(\OIIIHb) $<$ 1.1}), suggesting that the temperature and ionization parameter of the line-emitting gas are relatively uniform. The \NIIHa\ and \SIIHa\ ratios peak at the elongation and various regions in the cloud, suggesting that these regions may have high gas densities and/or significant shock excitation. 

There is clear spatial variation in the \OIIIHb\ ratios of component 2. The largest \OIIIHb\ ratios are observed in the vicinity of the central AGN and along the ionization cone. The \OIIIHb\ ratios decrease steadily with distance from the axis of the ionization cone, driven by a corresponding decrease in the number of ionizing photons. In contrast, the \NIIHa\ and \SIIHa\ ratios increase with distance from the axis of the ionization cone, consistent with our hypothesis that the observed anti-correlation between the diagnostic line ratios is driven by ionization parameter variations. 

We do not observe a strong relationship between the component 3 line ratios and position relative to the AGN ionization cone. The largest \NIIHa\ and \SIIHa\ ratios are found towards the north eastern side of the galaxy, in the vicinity of the elongation and the \NeIII\ peak. These large line ratios support our hypothesis that the gas in this region of the galaxy is primarily powered by fast shocks.

\begin{figure*}
\centerline{\includegraphics[scale=1,clip = true, trim = 0 190 0 0]{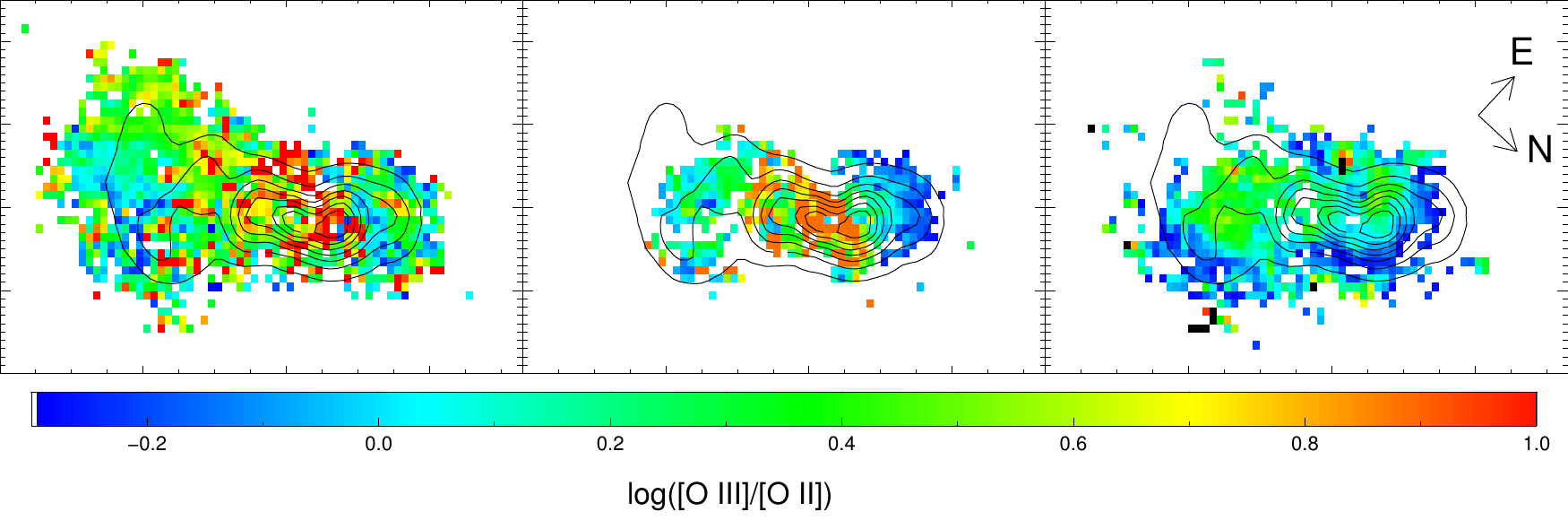}}
\caption{Maps of the \OIII/\OII\ ratios for each of the kinematic components. Contours trace the total \OIII\ emission. The \OIII/\OII\ ratio can be used to calculate the ionization parameter in \HII\ regions but is enhanced by collisional excitation in the presence of an AGN. In components 1 and 2 the ionization parameter decreases with distance from the axis of the AGN ionization cone, as expected. The average \OIII/\OII\ ratios in component 3 are significantly lower than those of the first two components, reflecting the prominence of shock excitation in ionizing the most turbulent gas in the system.}\label{Fig:ionization_param}
\end{figure*}

\subsection{Ionization parameter}
\label{subsec:ionization_param}
The ionization parameter $q$ is defined as the ratio between the flux of ionizing photons and the local hydrogen number density. This ratio is a measure of the maximum velocity at which the radiation field can drive an ionization front. The more commonly used dimensionless ionization parameter $U$ is related to $q$ by \mbox{$U = q/c$} (where $c$ is the speed of light). The combination of ionization parameter and electron density can be used to derive the photon flux (luminosity) required to ionize a particular region. In the following sections we explore the ionization parameter, electron temperature and electron density of the line emitting gas in J2240. Detailed calculations of the ionizing luminosity of the EELR are deferred to a future paper.

The ionization parameter in \HII\ regions can be calculated from the \OIII/\OII\ ratio given an estimate of the abundance of the line emitting gas \citep{KD02}. However, collisional excitation in the presence of an AGN increases the \OIII/\OII\ ratio for a given ionization parameter, making the conversion between \OIII/\OII\ ratio and ionization parameter difficult. Mapping the \OIII/\OII\ ratio can provide some insight into the spatial variation in the ionization parameter, although calculation of the absolute ionization parameter values is very uncertain.

We show maps of the \OIII/\OII\ ratios for each of the kinematic components in Figure \ref{Fig:ionization_param}. The \OIII/\OII\ ratio in component 1 peaks at the location of the \NeIII\ peak. To the south west of the AGN, the \OIII/\OII\ ratio is largest along the axis of the AGN ionization cone and decreases steadily with distance off the axis. This indicates that the number of ionizing photons decreases with increasing opening angle, consistent with expectations for an ionization cone.

Component 2 shows very similar \OIII/\OII\ ratios to component 1, consistent with our hypothesis that both components are ionized primarily by the central AGN. The peak \OIII/\OII\ ratios in component 2 occur in the vicinity of the central AGN, but large errors on the \OII\ flux values prevent robust calculation of the ratio over a significant fraction of the nuclear region. The average \OIII/\OII\ ratios in component 3 are significantly lower than those of the first two components. This is consistent with a larger contribution from shocks, which can have lower \OIII/\OII\ ratios than Seyfert 2 nuclei \citep[e.g.][]{Dopita95}.

The Dusty AGN models shown in Figure \ref{Fig:diagnostic_diagrams} can be used as a quantitative probe of the ionization parameter values across J2240. The line ratios of AGN-dominated spaxels ($\log$(\NIIHa) $< 0$) are best reproduced by models with \mbox{$\log U < -1.3$} at \mbox{$\alpha = 1.7$} or \mbox{$-2.6 < \log U < -1.3$} at \mbox{$\alpha = 1.4$}. The true $\alpha$ and $\log U$ values are likely to lie between these two estimates. The ionization parameter values we derive are larger on average than the values calculated by \citet{Schirmer13}. This is likely because we assume a metallicity of $\sim 2 Z_{\odot}$, where \citet{Schirmer13} base their calculations on the \citet{KD02} ionization parameter calibration assuming a metallicity of $\sim 0.5 Z_{\odot}$. 

\begin{figure*}
\centerline{\includegraphics[scale=1]{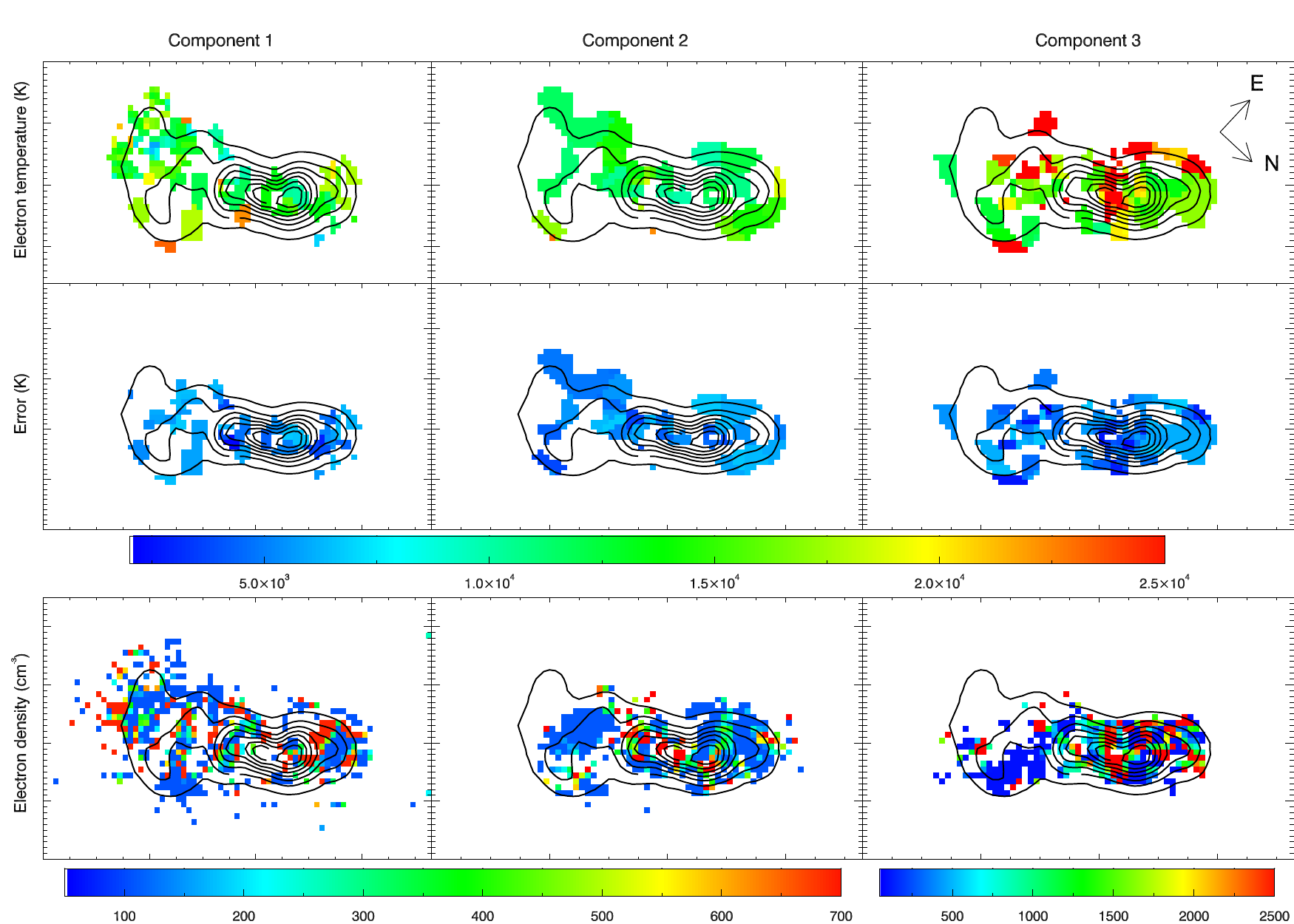}}
\caption{Maps of (top) electron temperature, (middle) error on the electron temperature and (bottom) electron density for each of the three kinematic components. Contours trace the total \OIII\ emission. Component 3 has the highest average electron temperature. The \NeIII\ peak, continuum peak and elongation all show electron temperatures between \mbox{20,000-25,000 K}, supporting our hypothesis that shock excitation is significant in these regions. Much of the diffuse gas in the cloud is in the low density limit, however, we are able to derive density measurements for the \OIII-bright regions along the AGN ionization cone, suggesting that material is being compressed along the ionization cone. The density reaches several hundreds to thousands of cm$^{-3}$ in the hottest, most energetic regions of the system.}\label{Fig:electron_temp}
\end{figure*}

\subsection{Electron temperature}
The temperature and density of electrons are also powerful probes of the principal ionization source(s) of the line-emitting gas. Compression and heating of gas along shock fronts can produce electron temperatures above \mbox{20,000 K} \citep[e.g.][]{Evans99}, compared to an average temperature of \mbox{$\sim$10,000-15,000 K} in AGN narrow line regions \citep[e.g.][]{Osterbrock06}. This provides a means to separate ionization due to fast shocks from ionization due to AGN activity.

The electron temperature determines the relative population rate for different excited states of the same atomic species. The \mbox{\OIIIc\ / (\OIIIa\ + \OIIIb)} ratio provides an estimate of the total cooling due to oxygen in the medium ionization zone of nebulae, and is therefore commonly used as an estimate of the electron temperature \citep{Osterbrock06}. \OIIIa\ and \OIIIb\ have excitation temperatures of \mbox{29,130 K} whereas \OIIIc\ is an auroral line with an excitation temperature of \mbox{62,094 K}. Therefore, the intensity ratio of the auroral line to the two oxygen strong lines increases with temperature. 

The relationship between the measured \OIII\ line ratio and the electron kinetic temperature depends on both the electron density and the intrinsic electron temperature distribution. The rate of collisional excitation increases with density, and therefore the \OIII\ ratio will be larger in high density nebulae than in low density nebulae at a given electron temperature. However, at densities below $n_e = 10^5 \rm cm^{-3}$ this effect is negligible and the \OIII\ ratio becomes independent of density \citep{Nicholls13}. \citet{Schirmer13} find that the electron density in J2240 rarely exceeds $650 \rm cm^{-3}$, indicating that electron density does not need to be considered in our electron temperature calculations. 

Traditionally, electron temperature calibrations have been based on the assumption that electron energies follow a Maxwell-Boltzmann distribution. However, \citet{Nicholls12} and \citet{Nicholls13} showed that a $\kappa$ distribution is more appropriate for \HII\ regions in the local universe. The $\kappa$ value controls the strength of the high energy electron tail. A $\kappa$ value of infinity corresponds to the Maxwell-Boltzmann distribution, and the fraction of hot electrons increases as the $\kappa$ value decreases. A larger fraction of hot electrons will increase the expected \OIII\ ratio relative to the Maxwell-Boltzmann estimate for a given kinetic temperature, causing calibrations based on the Maxwell-Boltzmann distribution to over-estimate the true nebular electron temperature. \citet{Nicholls12} show that correcting the electron temperature calibrations for the increased hot electron fraction appropriate to a $\kappa$-distribution resolves the long standing discrepancy between electron temperatures calculated using direct and strong line ratio methods. The correction is as follows \citep[see their Equation 36]{Nicholls13}:

\begin{equation}
T_{kin} = a(\kappa) T_{e,MB} + b(\kappa)
\end{equation}

where $T_{e,MB}$ is the electron temperature estimate according to the original Maxwell-Boltzmann calibration of \citet{Osterbrock06} and $a(\kappa)$ and $b(\kappa)$ are coefficients which depend on $\kappa$. We adopt $a$ = 14192.273 and $b$ = -2413.5652, calculated using Equations 37 and 38 of \citet{Nicholls13} with \mbox{$\kappa$ = 20}. The average $\kappa$ values of AGN narrow line regions are expected to lie between \mbox{$\kappa$ = 10-20}. Zones of very low $\kappa$ values (adjacent to regions of very high excitation) mix with equilibrium zones, producing intermediate average $\kappa$ values (David Nicholls, private communication). 

\citet{Schirmer13} analysed the electron temperatures in J2240 using the \OIII\ $T_e$ proxy, but due to the weakness of the \OIIIc\ auroral line in their data they were only able to derive globally averaged temperatures for the cloud and main body of the galaxy. They found the main body of the galaxy to have a higher average temperature than the cloud (\mbox{17,100 K} compared to \mbox{14,700 K}). However, such a globally averaged analysis masks any smaller scale temperature variations which would trace the localized impact of shock fronts on the energetics of the gas. Our GMOS data are much deeper and allow us to not only map the electron temperature across the galaxy, but also to do this on a component by component basis.   

Unfortunately, we are unable to detect \OIIIc\ at the 3$\sigma$ level in every component of every spaxel. We sacrifice a small amount of our spatial resolution by binning spaxels to achieve a target signal-to-noise ratio of 3. Traditional binning techniques decrease the spatial resolution over all regions of the galaxy uniformly, regardless of variations in the initial signal-to-noise ratio across the galaxy. Voronoi binning is an adaptive spatial binning technique which produces the optimal binning solution by tessellating bins of varying size \citep{Cappellari03}. Spaxels whose original signal-to-noise ratios exceed the target are not binned, and the bin sizes increase towards lower signal-to-noise regions. 

Figure \ref{Fig:electron_temp} shows electron temperature maps for each of the kinematic components. Component 3 has the highest average electron temperature, consistent with our identification of turbulent, shock- and AGN-ionized gas in this component. The \NeIII\ peak, the elongation and the continuum peak all have temperatures between \mbox{20,000 K} and \mbox{25,000 K}, consistent with our hypothesis that these regions are excited by a combination of shock excitation and AGN activity. 

The first two components, by contrast, are significantly cooler. The gas in component 1 has an average temperature of \mbox{14,700 K}, and the gas in component 2 has an average temperature of \mbox{12,900 K}, consistent with expectations for AGN narrow line region gas. Grouping the temperature measurements by physical region, we find that the average electron temperatures in the cloud and the main body of the galaxy are \mbox{13,700 K} and \mbox{13,900 K} respectively. We measure an \mbox{800 K} temperature difference between the two components, which is much smaller than the \mbox{2,400 K} difference found by \citet{Schirmer13}. However, if we include the more energetic third component in our calculations then the temperature of the main body of the galaxy rises to \mbox{15,500 K} and the temperature difference between the regions rises to \mbox{1,600 K}. 

The temperatures we derive for each of the components are systematically \mbox{$\sim$ 1,500 K} lower than the measurements of \citet{Schirmer13}. This is most likely to arise due to a combination of our multiple-component parametric emission-line fitting and our assumption of $\kappa$-distributed electrons. We check the consistency of our calculations by measuring the electron temperatures from the total \OIIIa, \OIIIb\ and \OIIIc\ fluxes in each spaxel (integrated across all three kinematic components), assuming a standard Maxwell-Boltzmann electron temperature distribution. We find that the cloud and the main body of the galaxy have average electron temperatures of \mbox{15,600 K} and \mbox{16,900 K} respectively, much closer to the values published by \citet{Schirmer13}. 

Correcting for extinction based on an unreddened Balmer decrement of 3.1 results in systematically lower electron temperatures across the galaxy. The component 1, 2 and 3 temperatures decrease by an average of \mbox{250 K} (\mbox{1.6 $\pm$ 0.4 per cent}), \mbox{195 K} (\mbox{1.5 $\pm$ 0.3 per cent}) and \mbox{480 K} (\mbox{2.2 $\pm$ 0.8 per cent}) respectively. These differences are a very small percentage of the total electron temperatures and therefore the choice of unreddened Balmer decrement does not have a significant impact on our electron temperature measurements.

\begin{figure*}
\centerline{\includemovie[toolbar, 3Dviews2 = views2_movie15.txt, text={\includegraphics[scale=1, clip = true, trim = 0 150 0 0]{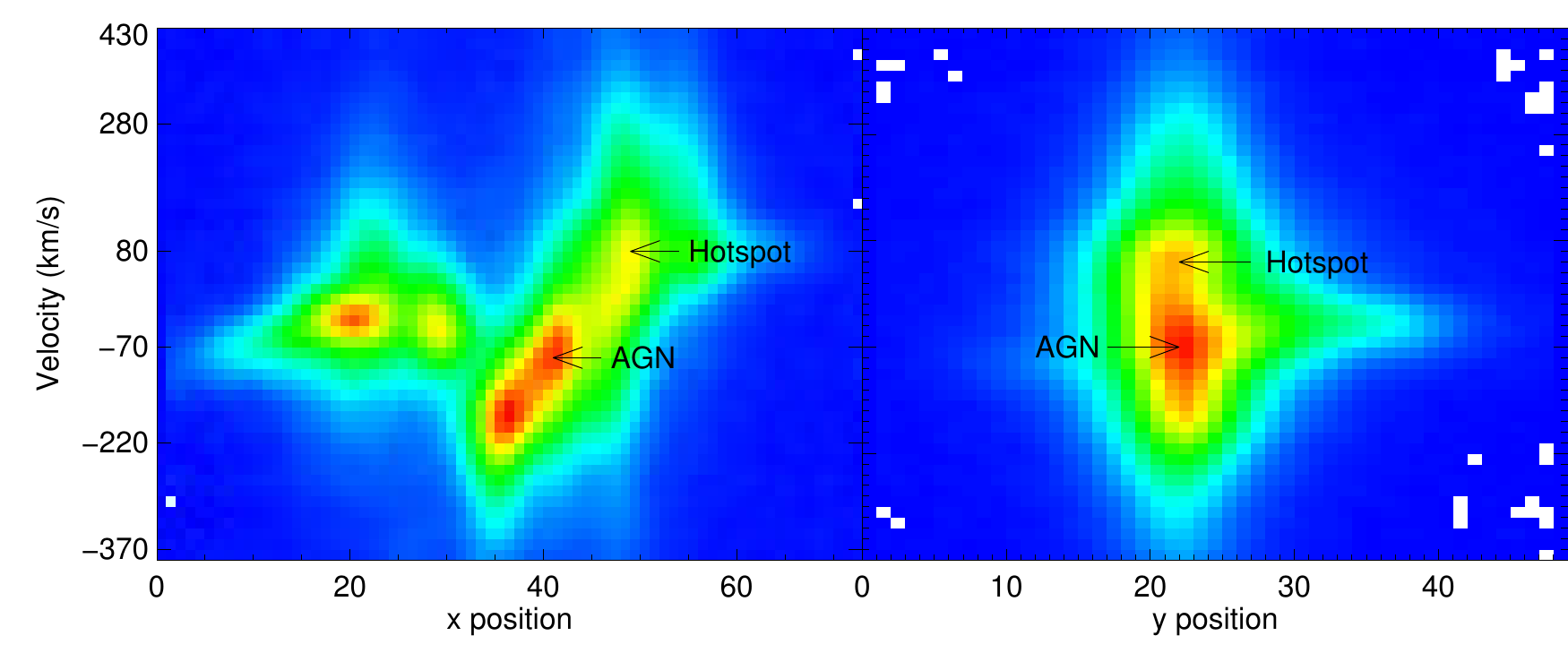}} ]{}{}{3dfig.u3d}}
\caption{Interactive 3D contour plot, produced in .wrl format using the \textsc{Python} package \textsc{Mayavi2}, converted to .u3d format using \textsc{PDF3D ReportGen} and embedded in the PDF file using the \LaTeX\ \textsc{movie15} package. By default, a 2D place-holder figure is displayed, but interactivity can be enabled in supported PDF readers by clicking on the figure. The interactivity allows the reader to rotate, pan and zoom in/out of the 3D plot. The plot shows iso-intensity surfaces for an 800 $\rm km \, s^{-1}$ slice of the reduced data cube extracted around the \OIIIa\ emission line. The red axis is velocity, going from \mbox{-420 $\rm km \, s^{-1}$} to \mbox{380 $\rm km \, s^{-1}$}, the green axis is Y position and the blue axis is X position. There are three pre-set views: position-position, velocity - \mbox{Y position} and \mbox{X position} - velocity, which can be selected from the ``Views'' drop-down menu. Two spheres embedded in the isosurfaces mark the position of the central AGN and the hotspot energized by interaction of the radio jet with the interstellar medium. These spheres are labelled but the text is not visible from all viewing angles.}
\label{fig:3D}
\end{figure*}

\subsection{Electron density}
\label{subsec:electron_dens}

The \SII$\lambda$6717/\SII$\lambda$6731 ratio is a probe of the electron density in the line-emitting gas. The two \SII\ lines result from transitions with similar ionization energies, however, the ratio between the maximum occupancies of the transitions is 1.5 and the ratio between their lifetimes is $\sim\frac{1}{3}$. In low density nebulae ($\rm n_e \la 100 \, cm^{-2}$), collisional excitation is very rare and the ratio between the transition strengths is approximately equivalent to the ratio of their maximum occupancies. At very high densities ($\rm n_e \ga 10,000 \, cm^{-2}$), the ratio of the transition strengths is approximately equal to the ratio of the maximum occupancies divided by the ratio of the lifetimes (0.44). However, at intermediate densities the probability of the \SII$\lambda$6717 transition increases relative to the probability of the \SII$\lambda$6731 transition (see \citealt{Osterbrock06} for a more detailed explanation). The \SII\ ratio ($R_{[S II]}$) also depends on temperature, according to the equation
\begin{align}
R_{[S II]} = a \frac{1 + bx}{1 + cx}
\end{align}
where $a$, $b$ and $c$ are constants and $x = 10^{-2} N T^{-0.5}$. We use the electron temperatures derived in the previous section as inputs to our electron density calculations. Because the \SII\ doublet is not sensitive to electron density for $n_e < 100 \rm \, cm^{-3}$, we present only an upper limit of $n_e < 100 \rm \, cm^{-3}$ for spaxels in which the estimated electron density is $\leq 100 \rm \, cm^{-3}$.

We show maps of the electron density for each of the kinematic components in the bottom row of Figure \ref{Fig:electron_temp}. We cannot calculate errors on the electron densities for any of the components due to the presence of skyline residuals in the region of the \SII\ line profile (see discussion in Section \ref{subsubsec:schematic}). 

The density maps highlight similar features to the electron temperatures maps. The average electron density in component 1 is \mbox{220 $\rm cm^{-3}$}, but this is a strict upper limit given that 60 per cent of the spaxels have only an upper limit on the electron density. The majority of the diffuse gas in the cloud is in this low density limit. However, we derive density measurements for the northern and southern \OIII\ peaks in the cloud, a clump of gas between the cloud and the main body of the galaxy, the \NeIII\ peak and the elongation on the north eastern side of the system, which are all expected to be dense regions given their temperatures, velocities and line ratios.

The average electron density in component 2 is also \mbox{220 $\rm cm^{-3}$}, but we again find that a large fraction of the spaxels (65 per cent) are in the low density limit and therefore the actual average density will be lower. The density clearly peaks along the axis of the AGN ionization cone, reaching \mbox{$n_e \, \sim$ 700 $\rm cm^{-3}$}. This is consistent with the findings of \citet{Schirmer13} and suggests that AGN activity is contributing to the compression of material. 

The average electron density in component 3 is much larger, reaching into the thousands due to the presence of some highly compressed gas. We observe that the densest gas in J2240 is also the hottest and the most energetic; indicating that all of the tracers and diagnostics we have explored paint a coherent picture of this system.

\section{Discussion}
\label{sec:discussion}

\subsection{Piecing together the puzzle: the 3D ionization structure of J2240}
Our analysis of the 3D kinematic and ionization structure of J2240 has confirmed that this system is very complex. Two narrow components and an underlying broad component are ubiquitous across the majority of the system, indicating that at least two physical gas components and at least two distinct ionizing sources contribute significantly to the emission signature of J2240. In the following section, we synthesise our results from Sections \ref{sec:initial_analysis} and \ref{sec:3D_analysis} to piece together a picture of J2240. 

Our discussion is complemented by an interactive 3D contour plot designed to assist the reader in their understanding of the velocity structure of J2240 (see Figure \ref{fig:3D}). The plot was produced in .wrl format using the \textsc{Python} module \textsc{Mayavi2}\footnote{The full documentation for the \textsc{Mayavi2} package is available at \href{http://docs.enthought.com/mayavi/mayavi/}{http://docs.enthought.com/mayavi/mayavi/}.} \citep{Ramachandran11}, converted to .u3d format using the proprietary software \textsc{PDF3D ReportGen}\footnote{The documentation for this software is available at \href{http://www.pdf3d.com/pdf3d\textunderscore reportgen.php}{http://www.pdf3d.com/pdf3d \textunderscore reportgen.php}}, and embedded in the PDF file using the \LaTeX\ \textsc{movie15}\footnote{\href{http://ctan.org/pkg/movie15}{http://ctan.org/pkg/movie15}} package. By default, a 2D place-holder figure is displayed, but interactivity can be enabled in supported PDF readers by clicking on the figure. The interactivity allows the reader to rotate, pan and zoom in/out of the 3D plot. The 3D plot shows iso-intensity surfaces for an \mbox{800 $\rm km \, s^{-1}$} slice of the reduced data cube extracted around the \OIIIa\ emission line. The red axis is velocity, going from \mbox{-420 $\rm km \, s^{-1}$} to \mbox{380 $\rm km \, s^{-1}$}, the green axis is Y position and the blue axis is X position. There are three pre-set views: position-position, velocity - \mbox{Y position} and \mbox{X position} - velocity, which can be selected from the ``Views'' drop-down menu. Two spheres embedded in the isosurfaces mark the position of the central AGN and the hotspot (discussed at the end of this section). These spheres are labelled, although the text is not visible from some viewing angles.

The 2D place-holder figure shows two position-velocity maps, constructed by summing the data cube slice over the $y$- (left) and $x$- (right) axes. The ionized gas cloud occupies a very restricted region of velocity space, compared to the main body of the galaxy which has a clear rotation curve with a velocity amplitude of \mbox{$\sim$300 $\rm km \, s^{-1}$}. The velocity of the gas cloud is very similar to that of the central AGN, consistent with the velocity fields shown in Figure \ref{Fig:vel_maps}. The clear morphological and kinematic association between the gas cloud and the main body of the galaxy indicates that there is a strong link between these two physical components of J2240. 

The line-emitting gas in the cloud typically has line ratios and electron temperatures consistent with ionization due to AGN activity. The \OIII\ peaks have large electron densities (\mbox{$n_e \, \sim$125 $\rm cm^{-3}$}) compared to the underlying gas in the cloud (\mbox{$n_e \, <$ 50 $\rm cm^{-3}$}); suggesting that gas may be compressed at shock fronts along the AGN ionization cone. The northern \OIII\ peak has the largest velocity dispersion in the cloud (clear from the large range of gas velocities visible at this spatial location; see the left hand panel of the 2D figure and the \mbox{X position - velocity} view of the 3D figure). The high velocity dispersion combined with strong \NeIII\ flux and large \NIIHa\ and \SIIHa\ ratios suggests that this region is strongly shock excited. The two southern \OIII\ peaks have relatively small velocity dispersions, large \OIIIHb\ ratios and small \NIIHa\ and \SIIHa\ ratios, indicating that the AGN is the dominant ionizing source in these regions. However, the detection of \NeIII\ and \HeII\ emission combined with electron temperatures of \mbox{$\sim$20,000 K} (especially in the peak closest to the central AGN) suggests that shock excitation is likely to contribute to the energy budget of these \OIII-bright regions. The low velocity dispersions are consistent with shocks propagating mainly perpendicular to the line of sight, along the AGN ionization cone.

The overlap region between the cloud and the main body of the galaxy is characterised by a superposition of the disc and cloud rotation curves. Double peaked emission line profiles are observed but restricted to a small region in which the emission from both components is of approximately equal intensity (see Figure \ref{Fig:narrow_comps}).

The main body of the galaxy has a compound emission signature which reveals both a narrow-line rotating component and a prominent broad component which is ubiquitous across the main body of the galaxy. Strong \NeV\ emission signposts the location of the AGN at the centre of the galaxy disc. The narrow emission component has similar line ratios and electron temperature to the line-emitting gas in the cloud, indicating that both are ionized primarily by the central AGN. The majority of the underlying broad component is also likely to be energized by the central AGN, whilst localised regions of fast shocks can be identified by their large \NIIHa\ and \SIIHa\ ratios, strong \NeIII\ and \HeII\ flux and large velocity dispersions. 

The elongation on the north eastern side of the galaxy is seen in both continuum and line emission, indicating that J2240 may have been morphologically disrupted during past interactions with a companion galaxy. Strongly redshifted, turbulent gas suggests the presence of outflowing material. Large \OIIIHb, \NIIHa\ and \SIIHa\ ratios and an electron temperature of \mbox{$\sim$25,000 K} provide additional evidence for significant shock excitation in the region. 

The \NeIII\ peak on north eastern side of the galaxy disc (see Section \ref{subsec:AGN_lines}) is a particularly interesting feature of J2240. It has a much larger \OIII\ flux than the neighbouring regions of the galaxy, and its \HeII\ flux is comparable to that of the central AGN. It has a peak velocity dispersion of \mbox{500 $\rm km \, s^{-1}$}, an electron temperature of \mbox{$\sim$ 20,000 K}, and similar line ratios to the AGN. The strong warp in the galaxy's \OIII\ emission profile appears to be driven primarily by strong emission from this concentrated region of turbulent, hot, dense gas. This hot spot may be energized by an AGN-driven jet or outflow interacting with the interstellar medium. J2240 has a \mbox{1.4 GHz} luminosity of $1.08 \times 10^{24} \rm \, W \, Hz^{-1}$, corresponding to a radio power of 6.8 $\times 10^{38} \rm \, erg \, s^{-1}$ and a jet power of $\sim \, 10^{42} \rm \, erg \, s^{-1}$ (using the $\rm P_{jet}-P_{1.4 GHz}$ scaling relation from \citealt{Cavagnolo10}). The localized nature of the shocked spot suggests that the interacting material is highly collimated, and therefore the hotspot is unlikely to be energised by an outflow alone. High resolution Very Long Baseline Interferometry (VLBI) observations are required to resolve and further analyse the radio structure of this target. 

Although each of the features we have discussed have distinct physical properties and are clearly resolved spectrally, the minimum separations between the features are 1-2 resolution elements (where one resolution element is the FWHM of the PSF) and therefore they are not spatially resolved. This does not impact our conclusions regarding the ionization sources and kinematics of J2240. Our 3D spectroscopic analysis has indicated that decomposing the spectra into different kinematic components allows us to spectrally resolve distinct ionizing sources, even when they are superimposed along the line of sight and therefore cannot be spatially resolved. 

\subsection{Fraction of \OIII\ luminosity due to AGN activity}
The large \OIII\ luminosity of J2240 makes it a very unusual object. The \OIII\ luminosity is both an order of magnitude larger than expected from its 22$\mu$m luminosity and an order of magnitude larger than that of typical Type-2 quasars at similar redshifts. However, our spectroscopic analysis has indicated that shock excitation is a non-negligible ionization source which can account for some of this discrepancy. We estimate the relative contribution of shock excitation to the \OIII\ luminosity of J2240 by assuming that shock excitation is responsible for all component 2 emission in the cloud region of the galaxy and all component 3 emission to the north east of the AGN. This is a conservative estimate given that the AGN narrow line region extends across the entire \OIII-emitting region of the system. We find that the AGN is responsible for at least 82 per cent of the galaxy's total \OIII\ luminosity (this number rises to 86 per cent when correcting for extinction using an unreddened Balmer decrement of 3.1). The contribution from shocks is clearly non-negligible and reduces the discrepancy between the \OIII\ and 22$\mu$m luminosities by a factor of $\sim$1.2, but is not large enough to account for the full order of magnitude discrepancy. This confirms that J2240 is a very unusual object, illuminated by the radiation field of an AGN which appears to have drastically reduced its accretion rate over less than the light crossing time from the central engine to the narrow line region.

\subsection{Origin of the Extended Emission Line Region around J2240: remnant of a quasar-driven outflow?}
\label{subsec:EELR}
To date, the majority of EELRs have been observed around three main classes of galaxies: the most powerful radio loud quasars at intermediate redshifts (\mbox{0.15 \textless\ z \textless 0.4}; \citealt{Stockton87,Fu09}), low redshift galaxies with weak central AGN (\mbox{z \textless\ 0.1}; \citealt{Keel12a}, \citeyear{Keel14}) and radio quiet luminous obscured quasars at \mbox{z $<$ 0.8} \citep{Greene11, Liu14, Hainline14}. The EELRs around the first two classes of galaxies extend to tens of kpc and are morphologically distinct from their host galaxies, whereas ionized gas around radio quiet luminous obscured quasars is ubiquitous to \mbox{$\la$ 10 kpc} but rarely extends beyond this radius, and appears to be morphologically associated with the host galaxy. This dichotomy suggests that the largest EELRs are either tidal in origin or outflow remnants \citep{Fu09, Keel14}, whilst smaller EELRs are limited by the availability of gas to ionize within the host galaxy \citep{Greene11, Liu14, Hainline14}. The EELR of J2240 extends over at least \mbox{22 x 35 kpc} \citep{Schirmer13} and is morphologically distinct from the host galaxy, and therefore clearly falls into the first group of EELRs.

The EELRs around intermediate redshift radio-loud quasars are thought to be remnants of quasar-driven superwinds triggered by galaxy interactions \citep{Fu09}. These regions are ionized by current strong AGN activity. The EELRs are morphologically distinct from their central galaxies and many contain bright, dense clumps of gas illuminated by shocks propagating through the gas. The gas clouds have high velocities but low velocity dispersions, and appear to be kinematically ordered on small scales but disordered on large scales \citep{Fu09}. In contrast, EELRs around low redshift galaxies with weak central AGN are thought to be tidal debris illuminated by AGN ionization echoes. The clouds are often diffuse and filamentary, resembling tidal tails or radial shell/ring structures. The gas clouds have line ratios consistent with AGN ionization, but otherwise appear to be kinematically very stable, following the ordered rotation curves of the galaxies they surround \citep{Keel14}. The objects in the \citet{Keel14} sample were selected to have at least a factor of 3 discrepancy between the observed AGN power and the AGN power required to illuminate the gas in the galaxy, and therefore provide an interesting comparison sample for GBs.

The morphology, ionization properties and kinematics of J2240 indicate that it is either a different type of object or at a different evolutionary stage to the objects in both the \citet{Keel14} and \citet{Fu09} samples. J2240 is neither radio loud nor a quasar host galaxy, but drives a radio jet with a synchrotron power of $\sim 10^{42} \, \rm erg \, s^{-1}$ which interacts with the interstellar medium of the galaxy and produces a bright, dense, turbulent hotspot. The \OIII\ luminosity of the EELR exceeds that of the most luminous quasars at the same redshift. This indicates that the quasar phase of J2240 must have ended more recently than the light crossing time from the central engine to the EELR ($\sim \, 10^4-10^5$ years; \citealt{Schirmer13}). We find clear evidence for hot, dense, luminous knots of gas along the ionization cone of the AGN. There is clear evidence for morphological disturbance in both the stellar continuum and ionized gas emission of J2240, consistent with the hypothesis that merger activity is a defining feature of systems with prominent EELRs. The entire galaxy appears to be embedded in a very low surface brightness asymmetric halo (see Figure 8 of \citealt{Schirmer13}), but the morphology of the ionized gas cloud suggests that it is unlikely to be composed of tidal debris. The morphology and \OIII\ luminosity of the EELR therefore suggest that J2240 may be a direct descendent of the luminous quasars observed in the \citet{Fu09} sample. 

The kinematic properties of J2240, however, point towards a more complicated evolutionary sequence, as noted by \citet{Schirmer13}. The EELR gas clouds appear to follow the large-scale rotation of J2240, making their motion globally ordered. The velocity structure of the line profiles in the south western region of the galaxy suggest that the EELR gas clouds have low velocities with high velocity dispersions. However, our multiple component analysis has indicated that the average velocity dispersion of the EELR gas is around 100 $\rm km \, s^{-1}$, and that the higher velocity dispersion components identified by \citet{Schirmer13} originate from gas associated with the galaxy disc which is superimposed in front of the cloud along the line of sight. We do not observe any high velocity gas in the cloud, perhaps because the gas in the cloud has very little motion in the radial direction or because the high velocity components are very faint and therefore undetected. The overall kinematic structure of the EELR in J2240 appears to be approximately intermediate between that of the \citet{Fu09} and \citet{Keel14} samples. 

Our observations of J2240 strongly suggest that its EELR is the remnant of a quasar-driven outflow rather than being tidal in origin. One possible evolutionary scenario is as follows. Strong gravitational interaction with a companion galaxy funnelled large amounts of cold gas towards the centre of the system, triggering a powerful quasar phase and simultaneously launching a radio jet and a quasar-driven outflow. A collimated outflow would naturally produce outflowing material with a coherent velocity structure and a strong morphological link to the central galaxy. The \OIII-bright knots along the AGN ionization cone are consistent with compression of outflowing material along fast shock fronts. The ordered rotation of the gas in the cloud is consistent with its close proximity to the central galaxy. A sustained quasar driven outflow would have the ability to remove the majority of the cold gas from the galaxy, starving the central engine and perhaps contributing to the rapid decrease in the accretion rate of the AGN. Molecular gas observations are required to further constrain the mass, kinematics and spatial distribution of molecular gas across the entire system.

It is worth noting that J2240 is significantly different from Hanny's Voorwerp. The ionization parameter in the gas surrounding the central engine of \mbox{IC 2497} is much lower than in J2240 (\mbox{$\log U \sim -3.2$} compared to \mbox{$\log U > -2.6$}). The highest measured velocities in the Voorwerp are on the order of $\sim 90 \rm \, km \, s^{-1}$ compared to J2240 where we measure velocities up to 400 $\rm km \, s^{-1}$ in the EELR and up to 600 $\rm km \, s^{-1}$ in the main body of the galaxy. The average electron temperature in the Voorwerp is \mbox{$\sim$ 13,500 K} but temperatures as large as \mbox{25,000 K} are measured in J2240. We also detect strong \HeII\ emission in many \OIII-bright clumps offset from the centre of J2240, which is not observed in the Voorwerp. All of these differences indicate that shock excitation is much more prominent in J2240 than in the \mbox{Voorwerp - IC 2497} system. However, both systems show evidence for the onset of outflows corresponding closely to the rapid decrease in brightness of their central engines; suggesting that the evolutionary scenarios of the two systems may not be significantly different. 

Whilst the EELR of J2240 appears to be the remnant of a quasar-driven outflow, the GB selection criteria do not guarantee the presence of gas \mbox{$>$ 10 kpc} from the nucleus of the host galaxy. The angular extents of the 29 GBs presented by \citet{Schirmer13} imply physical Petrosian radii of \mbox{4.5 - 15 kpc}, and 71\% of the sample have Petrosian radii \mbox{$<$ 10 kpc}, suggesting that gas reservoirs extending \mbox{$>$ 10 kpc} may not be a defining characteristic of the GB phenomenon. Follow up observations of a larger sample of GBs are required to determine whether or not their EELRs have a common origin, and whether differences in the physical properties of EELRs are driven primarily by intrinsic properties of the host galaxy or by the mechanism which triggers the rapid decrease in accretion rate of the central engine. 

\subsection{Space density of quasar ionization echoes}
The space density of GBs (4.4 $\rm Gpc^{-3}$) is very low, suggesting that the progenitors of GBs must be extremely rare. We use the redshift-dependent quasar luminosity function derived by \citet{Ross13} to investigate the expected space density of galaxies representing a $\sim 10^5$ year phase in the lifetimes of luminous quasars. We convert the \OIII\ luminosities of the GBs into approximate rest-frame $i$-band magnitudes using the conversion given by \citet{Reyes08}. All GBs in our sample have \mbox{$M_i < -25$}. We integrate the quasar luminosity function from \mbox{$M_i$ = -25} to \mbox{$M_i$ = -$\infty$} at both \mbox{z = 0.19} and \mbox{z = 0.35}, the boundaries of the redshift window for the lower redshift GB sample. We find that the expected space densities of luminous quasars at \mbox{z = 0.19} and \mbox{z = 0.35} are \mbox{$\sim$ 1300 $\rm Gpc^{-3}$} and \mbox{$\sim$ 3300 $\rm Gpc^{-3}$} respectively. We then use a simple timescale argument to determine the expected ratio of quasars to GBs. We assume that the lifetimes of luminous quasars are approximately equivalent to the Salpeter time (the $e$-folding time for Eddington-limited black hole growth; \citealt{Salpeter64}) which is 4.2 $\times$ $10^7$ years (see e.g. \citealt{Hopkins09}). We also assume that the average time required for light to cross the EELRs of GBs is $10^5$ years \citep[e.g.][]{Schirmer13, Keel14, Keel12a, Lintott09}. We then multiply the derived quasar space densities by the ratio of the GB lifetime to the quasar lifetime to derive estimates for the expected space densities of GBs. We find expected space densities of 3.1 $\rm Gpc^{-3}$ and 7.9 $\rm Gpc^{-3}$ at \mbox{z = 0.19} and \mbox{z = 0.35} respectively. The observed space density of GBs lies in between these two numbers and is therefore consistent with a scenario in which GBs are a common but short-lived stage in the evolution of the most luminous quasars (\mbox{$M_i$ $<$ -25} at \mbox{0.3 $<$ z $<$ 0.7}). 

\section{Summary and Conclusions}
\label{sec:conclusions}
We have used Gemini-GMOS IFU data to conduct a detailed analysis of the 3D ionization state and kinematics of the GB galaxy J2240. The majority of the optical emission from this system is concentrated in strong, high equivalent width nebular emission lines (e.g. \OIIIa, \OIIIb, \Hb, \Ha, \NII, \OII, \SII) which overpower the underlying stellar continuum emission. The emission line profiles are very complex, indicating that several kinematic components contribute significantly to the observed emission signature of the galaxy. We use the parametric emission-line fitting code \textsc{lzifu} (Ho et al., in prep) to model the spectrum of each spaxel as a linear superposition of up to three Gaussian components with varying velocities and velocity dispersions and thus reconstruct the emission spectrum of each kinematic component individually. Our analysis of the \OIII\ morphology, velocity field, ionization state, and electron temperature and density of each of the kinematic components has indicated that each of the components have distinct physical properties:

\begin{itemize}
\item{Component 1 has the smallest average velocity dispersion (\mbox{$\sigma \, \sim \, 100 \, \rm km \, s^{-1}$}) and traces the bulk motion of ionized gas in the cloud. The majority of the cloud material lies at the systemic velocity of the galaxy and has a strong morphological and kinematic link with the galaxy's central engine. The remainder of the cloud material appears to be co-rotating with the main body of the galaxy but with a shallower rotation curve than that of the disc material. The strongest \OIII\ emission comes from clumps of gas which lie along the ionization cone of the central AGN. The \OIIIHb, \NIIHa\ and \SIIHa\ ratios of the gas in the cloud indicate that the AGN is the dominant ionizing source. The average electron temperature is \mbox{$T_e \sim$14,700 K}, consistent with previously measured values for AGN narrow line regions.}
\item{Component 2 has a slightly larger average velocity dispersion (\mbox{$\sigma \, \sim \, 150 \, \rm km \, s^{-1}$}) and traces the bulk motion of gas associated with the main body of the galaxy, as well as shocked material in the cloud. The large scale gas motions reveal a rotation curve with an amplitude of $v \sim \, 300 \, \rm km \, s^{-1}$ across the main body of the galaxy. The centre of this rotation curve is coincident with the peak of the galaxy's continuum emission which is also a \NeV\ source and is therefore likely to be the central AGN. The diagnostic line ratios indicate that the AGN is again the dominant ionizing source. However, \NeIII\ and \HeII\ detections in dense, turbulent clumps of gas in the cloud indicate that shock excitation also contributes significantly to the ionization signature of this component. The average electron temperature is \mbox{$T_e \sim$12,900 K}.}
\item{Component 3 traces very turbulent gas (\mbox{$\sigma \, \sim \, 300-800 \, \rm km \, s^{-1}$}) in the main body of the galaxy, ionized by fast shocks and the hard ionizing radiation field of the central AGN. The average electron temperature in this component is \mbox{$T_e \, \sim$ 19,300 K}. There is some evidence for ordered rotation, but there are many pockets of gas whose velocities deviate significantly from the underlying rotation curve. The diagnostic line ratios indicate that the majority of this component is likely to be excited by the hard ionizing radiation field of the AGN. However, \NeIII\ and \HeII\ detections in regions with large \NIIHa\ and \SIIHa\ ratios and high electron temperatures (\mbox{$T_e \, \sim$ 20,000 K}) again indicate that fast shocks are a prominent ionizing source in this component. We identify a turbulent hotspot to the north east of the AGN whose properties are consistent with a radio-jet shock-heating the interstellar medium.} 
\end{itemize}

Our analysis confirms that J2240 is a very unusual object, with disturbed morphology, complex kinematics and several ionizing sources. Our results are both qualitatively and quantitatively consistent with the findings of \citet{Schirmer13}, but provide a much more detailed insight into the principal ionizing sources of J2240, allowing us to estimate the contribution of the AGN to the total \OIII\ luminosity. We find that the AGN must be responsible for at least 82 per cent of the total \OIII\ luminosity of J2240. Shock excitation is a non-negligible ionizing source but is not significant enough to resolve the discrepancy between the \OIII\ and 22$\mu$m luminosities of J2240 (see \citealt{Schirmer13}). This confirms that the EELR in J2240 is a quasar ionization echo. 

The evolutionary history of J2240 remains a puzzle. Although the EELR of J2240 has slightly different properties to similar regions observed around radio-loud quasars, our observations indicate that the cloud may be the remnant of a quasar-driven outflow, triggered by previous a galaxy interaction and now illuminated by a quasar ionization echo. VLBI and molecular gas observations are required to constrain the morphology of the radio emission and the mass, kinematics and spatial distribution of molecular gas in the system. 3D spectroscopic analysis and deep imaging of a larger sample of GBs will reveal whether these objects share similar properties to J2240, and ultimately will allow us to determine the evolutionary history of the GB phenomenon.

\section{Acknowledgements}
The authors would like to thank the referee for their helpful comments which improved the clarity of this manuscript. RD acknowledges the support of an Australian Gemini Undergraduate Summer Studentship from the Australian Astronomical Observatory funded by Astronomy Australia Ltd. RD would like to thank Geoff Bicknell for valuable discussions regarding radio jets and AGN-driven outflows. Based on observations obtained at the Gemini Observatory (acquired through the Gemini Science Archive and processed using the Gemini IRAF package), which is operated by the Association of Universities for Research in Astronomy, Inc., under a cooperative agreement with the NSF on behalf of the Gemini partnership: the National Science Foundation (United States), the National Research Council (Canada), CONICYT (Chile), the Australian Research Council (Australia), Minist\'{e}rio da Ci\^{e}ncia, Tecnologia e Inova\c{c}\~{a}o (Brazil) and Ministerio de Ciencia, Tecnolog\'{i}a e Innovaci\'{o}n Productiva (Argentina). Ureka is a collaboration of Space Telescope Science Institute and Gemini Observatory. We thank Bryan Miller for his contributions to the \emph{ifudrgmos} package used for processing our data.
 
\bibliography{mybib}

\appendix

\section{Modifications to GMOS IRAF package}
We reduced our data using a modified version of the Gemini GMOS IRAF package, as described briefly in Section \ref{subsec:data_reduction}. Here we include a detailed description of our modifications.

\subsection{Bias and overscan}

To avoid errors due to charge-transfer contamination in the overscan region, the bias zero-level was determined with the \emph{gfreduce.biasrows} parameter set to ``3:64'' for GMOS-S and ``6:46'' for GMOS-N with the EEV deep-depletion (EEV-DD) CCDs, selecting a small, unilluminated section at one end of the overscan region.

\subsection{Bad columns}

For our GMOS-S data, we used \emph{gemarith} to create an integer bad pixel mask from a science exposure and \emph{imreplace} to flag pixels that correspond to obvious bad columns (and a small, associated glow region), based on inspection. Separate masks were produced for science, standard star and calibration exposures, as the presence or extent of some bad columns was observed to vary between exposures of different length and brightness. The masked pixel values were replaced with a low-order (5th--32nd, depending on the image structure) fit to each row (spectrum), using a new wrapper task, \emph{gemfix}, which has options to use IRAF's \emph{fit1d} or \emph{fixpix}.

For the GMOS-N EEV-DD detectors, no bad columns were apparent, so no bad pixel masks were applied prior to cosmic ray rejection.

\subsection{Scattered light}

Scattered light in the background of flat-field and scientific spectra was modelled by fitting a 5th- to 9th-order surface to the values of unilluminated image rows between blocks of fibres, over each CCD amplifier. This model was then subtracted from the data, providing an order-of-magnitude improvement in the nominally-unilluminated zero-levels. The contamination in question is present at the level of up to $\sim$15 per cent over the useful range of the data (deteriorating to $\sim$100 per cent at the faint extremes). This processing was achieved using the \emph{ifudrgmos} scripts \emph{fndblocks} and \emph{gfbkgsub}. Flat-field spectra were extracted both before and after this procedure, first to determine the fibre and gap locations, as input to \emph{fndblocks}, and then with the scattered light removed for more accurate results.

\subsection{Wavelength calibration}

The wavelength calibrations for our blue and green channels are derived from the day-time CuAr spectra provided by Gemini. Only the $\sim$20--40 brightest lines (with $>$100 counts) were used, minimizing any centroid errors due to charge trailing from faint pixels during readout, which we have observed in previous GMOS datasets (see \ \citealt{Goudfrooij06}).

For the green channel, each exposure's wavelength zero point was adjusted to account for possible flexure with respect to the arc, by measuring 13 lines from \citet{Osterbrock96} in the sky spectrum using \emph{rvidlines} and editing the FITS CRVAL1 keywords accordingly; this resulted in an RMS absolute correction of 0.5\AA\ and an RMS correction relative to the first exposure of 0.4\AA.

For the blue channel, the only sky line detected ([N~I] at 5199\AA) had too low a S/N in two exposures to provide accurate flexure correction, so none was applied, but our low-S/N measurements indicate an RMS difference between exposures of 0.3\AA\ ($<$0.1\AA\ between those with a strong detection), consistent with the moderate range of elevation (55--60\degree) and rotation (45\degree) of the observation.

For the red channel, the provided CuAr spectrum is very sparse, so 17 strong sky lines were instead measured directly in each object exposure and compared with \citet{Osterbrock96} to derive wavelength solutions. These solutions are consistent between fibres, unaffected by the much weaker object flux. Only two sky lines coincide with (still several times fainter) line emission from the source. Some ghost lines at the blue end of the red slit were ignored and do not overlap with our range of scientific interest.

\subsection{Flat fielding}

A normalized flat field spectrum was produced using a version of \emph{gfextract} that we have re-written previously to account for wavelength distortions (the original version divides spectra with slightly differing wavelength solutions by a single vector of fitted continuum values). This change avoids minor artificial roll-off in flux towards the corners of the stack of extracted spectra and, in particular, eliminates $\sim$5 per cent discontinuities between the two halves of the IFU field. When fitting the continuum, our algorithm produces a steep drop at each extreme of the spectrum, so we use the \emph{gfresponse.sample} parameter to restrict the fit to the useful wavelength range, plus the first and last pixels to constrain the ends.

Our flats were generated using only GCal quartz-halogen exposures, omitting the optional twilight exposures (\emph{gfresponse.skyimage}) that are often used to account for any slow turn-down in GCal intensity over the field. It is well known that gravitational flexure can lead to variations in the throughput of optical fibres (eg. \citealt{Poppett10}), depending on their individual stresses. When processing GMOS-IFU data using twilight exposures, which are taken at a different pointing from the target, we have encountered some number of artifically bright or dark spectra, consistent with flexure effects, which we do not see using only contemporaneous GCal flats. Moreover, the $5\times7$ arcsec IFU is only a fraction of the size of the 5.5 arcmin GMOS imaging field, over which GCal flatness is much more of an issue, and we do not observe measurable roll-off over the IFU field. Even when test-processing twilight spectra in place of scientific exposures, the spatial residuals do not exhibit a low-order trend at the detectable level of a few percent.

\subsection{Cosmic ray cleaning}

Cosmic rays were identified using the spectroscopic version of L.A.\ Cosmic for IRAF (via the \emph{gscrspec} wrapper script in \emph{ifudrgmos}), with a modification to the \emph{imcalc} step at line 277 to avoid artifacts in the presence of negative values. We then re-cleaned the applicable pixels using our new \emph{gemfix} script with its \emph{fixpix} option, providing slightly more accurate replacement values than L.A.\ Cosmic's running median. Pixels flagged as cosmic rays are also later excluded during spatial interpolation by \emph{gfcube}.

\subsection{Flux calibration}

The standards taken for this programme in photometric conditions were used for absolute spectrophotometric flux calibration. These were processed similarly to science exposures but with only light ($\sim10\sigma$) cosmic ray cleaning, which is sufficient to avoid a bias due to cosmic rays while avoiding spurious rejection caused by L.A.\ Cosmic not being optimized for the strong modulation in bright fibre IFU spectra.  The gap between the two brightest blocks of fibres was omitted when fitting scattered light for the blue channel, due to the strength of the PSF wings in that nominally-unilluminated region. For the red channel, the sky lines in the longer scientific exposure were used for wavelength calibration. A simple flux summation over the IFU object fibres (with \emph{gfapsum}) was used to extract the standard spectra. For the green channel, a non-photometric standard taken close in time to the science data was used to derive an initial response function, whose absolute level was then corrected using a standard obtained in photometric conditions 2 months later (both giving very close results). We confirm from our standard PSFs in reconstructed images that a negligible proportion of flux was lost outside the area of the IFU field, so we have not corrected for this.

Our final spectra were compared with tabulated fluxes provided in the \emph{gmos\$calib} (for EG~131) or \emph{onedstds\$spec50cal} IRAF directories, adding some interpolated bandpasses (with the flux in the surrounding bins varying by $\leq\!5$ per cent) where necessary to constrain the ends of the fit, mainly for the red channel. After applying the resulting sensitivity function to the science data, we used a new \emph{gfcube} option to convert the flux per fibre to units of $\rm erg\;cm^{-2}\;s^{-1}\;\AA^{-1}\;arcsec^{-2}$ in our data cubes.

No compensation was applied for slightly different variations in quantum efficiency with wavelength between the 3 individual GMOS CCDs, as no measurements are available to date for the GMOS-N EEV-DD detectors and the processing functionality is still under development by Gemini. For the GMOS-S CCDs we expect resulting discontinuities in flux at the level of $\sim$2 per cent at the chip gap wavelengths (B.~W.\ Miller, private communication), which are not obvious in the data and would not significantly alter our results.

\subsection{Reconstruction and combination of data cubes}
\label{subsec:cube_reconstruction}

Our calibrated spectra were resampled to 3-dimensional data cubes with \emph{gfcube}, using its built-in compensation for atmospheric dispersion (modelled using the SLALIB library), which we added previously for a different project. We have also added an option to propagate approximate variance alongside the science array and used the resulting error estimates in our analysis. These errors are approximate because of the statistical correlation between adjacent pixels generated by resampling IFU fibres onto a finer spatial grid in the data cubes. To avoid computationally-expensive propagation of covariance arrays, we simply upscale the variance by a factor of $\sqrt{12}$ (the ratio of spatial pixel densities) such that noise estimates will be correct when averaging over more than $\sim3$ pixels spatially.

The data cubes corresponding to individual exposures were combined using our aforementioned \emph{PyFU} Python scripts. We have added new options to facilitate registration of nebulous structure via spatial cross-correlation (\emph{pyfalign.method=``correlate''}) and to inspect the resulting alignment more easily (\emph{pyfmosaic.separate=yes}), as well as propagating variance automatically.

A new PyFU script, \emph{pyflogbin}, was used to resample the co-added data cubes (and their variance) in bins of constant $\Delta \lambda / \lambda$, for analysis, following the convention of \citet{Greisen06}, eq.~5; this is conveniently similar to the linear input WCS but differs from IRAF's representation. This resulted in constant radial velocity increments of 28, 22 and 11 $\rm km \, s^{-1}$ in the blue, green and red channels respectively.

\end{document}